\documentclass[%
nofootinbib,
 amsmath,amssymb,
 aps,
 prd,
 12pt,
tightenlines
]{revtex4-2}
\usepackage[utf8]{inputenc}
\usepackage{graphicx}
\usepackage{dcolumn}
\usepackage{bm}
\usepackage{xcolor}
\usepackage{comment}
\usepackage{amsmath}
\usepackage[english]{babel}
\usepackage[autostyle, english = american]{csquotes}
\MakeOuterQuote{"}
\usepackage{hyperref}

\newcommand{\val}[1]{{\color{red}{#1}}}

\begin{document}

\preprint{APS/123-QED}

\title{Analog regular black holes and black hole mimickers in a Bose-Einstein condensate: from the core to the light rings}

\author{Valentin Pomakov}
 \email{vpomakov@sissa.it}

\author{Stefano Liberati}%
 \email{liberati@sissa.it}
\affiliation{%
 SISSA, Via Bonomea 265, 34136 Trieste, Italy \\ INFN, Sezione di Trieste, Via Valerio 2, 34127 Trieste, Italy \\ Institute for Fundamental Physics of the Universe – IFPU, Via
Beirut 2, 34014 Trieste, Italy
}%

\date{\today}

\begin{abstract}
We show how static, spherically symmetric regular black holes and black-hole mimickers with radius-dependent mass function $m(r)$ can be emulated in a quasi-two-dimensional Bose--Einstein condensate, with the aim of probing inner-horizon and inner-light-ring instabilities in the laboratory. Matching the acoustic and gravitational line elements on an equatorial disc fixes the ratio of radial flow velocity to sound speed and reproduces both horizon locations and surface gravities exactly. The analogue geometry nevertheless contains a residual radially varying sound-speed factor in $g_{tt}$, which shifts the light-ring positions and reduces the range of the regularization parameter $\ell/M$ for which a light-ring pair exists. We derive the interaction profile that removes this shift. Combining the metric correspondence with the continuity and Bernoulli equations, we then obtain closed-form expressions for the external potential and interaction strength realizing the full target spacetime, with $M$ and $\ell$ directly tunable through parameters of the potential. Finally, we translate the construction into experimental requirements (quasi-two-dimensional confinement, healing-length resolution, three-body lifetime, and stationary-flow lifetime) and present three fiducial designs achievable with present-day $^{39}\mathrm{K}$ platforms.
\end{abstract}

\maketitle


\section{\label{sec:Intro}Introduction}
There has been renewed interest in exploring alternatives to the standard black hole solutions of general relativity thanks to the outstanding progress in observations through direct gravitational-wave detections and imaging with very long baseline interferometry. In particular, one such alternative are \emph{regular black hole} (RBH) solutions, whose interior is non-singular and described by a maximally symmetric geometry. These spacetimes occur either by modifying the r.h.s. or the l.h.s. of the Einstein field equations -- either the stress energy tensor is non-vanishing and describes, for example, a non-linear electrodynamics configuration \cite{Balart_2014}, or, equivalently, gravity is modified to include quantum-gravitational effects. Indeed, it is expected that a complete theory of quantum gravity would resolve the issue of singularities, considered to signal the breakdown of applicability of General Relativity (GR). The RBH paradigm is a phenomenological approach towards this goal in the absence of a fully consistent quantum gravity theory.

Another alternative to GR's singular black holes are \emph{black hole mimickers} (BHMs). These are horizonless stars that are compact enough to feature light rings much like black holes. Provided a mechanism different from an outer horizon to impede BHMs from radiating light away to observers at infinity, these objects are valid alternatives to the "black holes" as observed in images produced by the Event Horizon Telescope Collaboration \cite{2019_EHT1, 2019_EHT4, visser2009smalldarkheavyblack} via their light rings and their shadow. BHMs, apart from being valid contenders to black holes observationally, elegantly solve both the singularity problem and the information loss problem. The literature abounds in models of the matter distribution that would yield such objects (gravastars, fuzzballs, boson stars, topological stars, etc.) \cite{bambi2025blackholemimickerstheory}, and much like RBHs, they arise in a variety of semiclassical and quantum-gravity frameworks.

Non-rotating RBHs and BHMs, both contenders to GR's standard singular black holes, can be described by the same class of simple phenomenological static, spherically symmetric spacetimes that we will call by the overarching term \emph{regular ultracompact object} (regular UCO) spacetimes, sometimes leaving the regularity implicit. In addition to the anyway present Arnowitt–Deser–Misner (ADM) mass of the spacetime, their Misner--Sharp--Hernandez mass function $m(r, \ell)$ features a single new parameter: the regularization scale $\ell$, that regularizes curvature scalars at the center $r\to0$. The spacetime in question is

\begin{equation}
\label{eq:ST}
    ds^{2}_{\textsc{gr}} = - \left( 1-\frac{2m(r)}{r} \right)dt^2 +  \left(1-\frac{2m(r)}{r}\right)^{-1}dr^2 + r^2d\Omega^2
\end{equation}
where $d\Omega^2 = d\theta^2 + \rm{sin}^2\theta d\phi^2$. Depending on the value of $\ell$, the horizon condition $1-\frac{2m(r)}{r} = 0$ has a different number of real solutions, and therefore this one-parameter family of spacetimes can be a
\begin{enumerate}
    \item RBH, featuring a trapped region sandwiched between an outer and an inner horizon, with a regular, typically de-Sitter core inside the latter;
    \item BHM with no horizons at all, but \emph{at least two} light rings: the usual outer unstable null orbit and a geometrically stable inner one.
\end{enumerate}

Both of the novel features, the inner horizon in the case of RBHs, and the inner light ring in the case of BHMs, are plagued by instabilities.
The inner horizon is classically unstable to perturbations exhibiting exponential growth of the internal mass parameter and divergence of curvature variables near the horizon, as first demonstrated in \cite{PoissonIsrael1989} in the context of rotating and charged singular black holes, and later established rigorously for the Reissner--Nordstr\"om and Kerr cases \cite{Luk2015Proof, Moortel2019The, Burko2016Ja} and for quantum-corrected black holes \cite{Cao2023The}. Subsequent studies have shown that this instability persists in regular black hole models and under generic dynamical conditions. The reason for this is that this so-called \emph{mass inflation instability} plagues \emph{future Cauchy horizons} in general and it can be understood as follows. Due to the fact that the entire infinite history of the exterior region lies in the causal past of this null hypersurface, signals that reach it experience an exponential blueshift. Provided both an ingoing and an outgoing null flux of perturbations, the mutual concentration of exponentially blueshifted energy density at the inner horizon translates (via the Einstein field equations) into an exponential growth of the mass parameter with a timescale set by $1/c|\kappa_-|$, where $\kappa_-$ is the inner horizon surface gravity; see, e.g.~\cite{Carballo-Rubio:2018pmi, Carballo-Rubio:2021bpr, DiFilippo:2022qkl} and references therein. 


Remarkably, this instability persists even in \emph{dynamical} black holes with slowly evolving inner horizons. In this case, the exponential accumulation of energy occurs over a finite time interval, but can nevertheless be efficient enough to produce significant backreaction on the spacetime~\cite{Carballo-Rubio:2024dca,Cardoso_2023}. Even in configurations engineered to suppress classical mass inflation, such as ``inner-extremal'' black holes with $\kappa_-=0$~\cite{Carballo-Rubio:2022kad}, semiclassical backreaction from quantized fields can further modify the inner horizon region~\cite{McMaken_2023}. In particular, the renormalized stress--energy tensor of a scalar field in the Unruh vacuum can diverge whenever $|\kappa_-| \neq |\kappa_+|$~\cite{McMaken_2023,Balbinot:2023vcm}, where $\kappa_+$ is the outer horizon surface gravity. Thus, singular behavior arises even in stationary inner-extremal black holes. For dynamically formed black holes in the so-called ``in'' state, however, a milder, power-law instability is induced~\cite{Carballo-Rubio:2026gwg,Arrechea:2026rua}.

On the other hand, in the case of horizonless BHMs, null perturbations can accumulate close to the inner light ring, it being a geometrically stable circular orbit (a potential minimum). It is located between the potential barrier of the outer light ring and an inner, reflective structure of centrifugal nature. Null perturbations can become "trapped" in this structure for a long time, corresponding to long-lived modes in the QNM signal. For this accumulation to cause a dynamical instability and to backreact on the background spacetime, a sustained flux of perturbations is required, since no Bogolyubov particle generation mechanism is present. It has been shown that, in the eikonal limit, long-lived perturbation modes are controlled by the light-ring structure and indeed peak near the inner light ring. While such linear-perturbation results are considered a red herring of an instability at the non-linear level, a full proof of the inner light ring instability would require full numerical nonlinear analysis taking into account potential dissipative effects (\cite{Cardoso_2014, Franzin_2024, Cunha_2017, Cunha_2023}, see also \cite{Redondo-Yuste:2025hlv, Benomio:2024lev} for the closely related problem of turbulence in spacetimes with stable trapping). The number and nature of the light rings of such objects is discussed in \cite{hod2017numberlightringscurved, Di_Filippo_2024}, and their observational prospects in \cite{Carballo-Rubio:2022aed, Carballo_Rubio_2025}.

Analog gravity can serve as testbed for the inner horizon and the inner light ring instabilities. Analog systems can provide faithful representation of the linear, test-field regime of these instabilities and thus of their onset, which relies on phenomena of purely kinematic nature \cite{LivingReviews}. In contrast, analog systems are not representative of the fully nonlinear gravitational dynamics of backreaction, being governed by a different set of dynamical equations\footnote{Albeit this does not necessarily imply that there would not be a backreaction of the perturbation growth and that it would be necessarily lead to a different qualitative behaviour of the GR one.}. Another advantage of the analog-gravity approach is the possibility of probing the ultraviolet-robustness of these instabilities, the inner horizon one in particular being determined by an exponential blueshift. More concretely, the BEC analog platform, for instance, exhibits dispersive deviations from relativistic wave propagation at sufficiently high frequencies that mimic Lorentz-breaking quantum-gravity effects expected in an effective field theory framework, a noteworthy example being Hořava gravity \cite{Herrero-Valea:2023zex}.

In the horizonless case, regardless of the dispersion relation, a key observational signature are echoes: the time-domain counterpart of the long-lived perturbations trapped close to the inner light ring. The time delay between echoes is closely tied to compactness, which depends on the regularization scale $\ell$ in the spacetimes of interest here \cite{Pomakov:2026alh}. In an analog-gravity setup where $\ell$ can be tuned, one could tune the echo time delay and potentially the timescale for onset of the inner light ring instability, making it natural to seek an experimental analogue of echoes. Of course, the echo signal is closely tied to the presence of the inner light ring, and therefore carries a reliable signature of the onset of the instability. If the object evolves rapidly towards a new metastable equilibrium (like a black hole \cite{Carballo_Rubio_2018} or a less compact object \cite{Cardoso_2014}), the echo signal might be largely deformed or disappear \cite{Vellucci_2023}, which is another aspect that analog-gravity could provide hints for. For a more detailed summary of relevant RBH and BHM phenomenology, the reader is referred to the present authors' work on a similar program relying on surface-gravity waves in fluid surfaces as analog platform \cite{Pomakov:2026alh}.

To realize this program, we need to make sure that, in the analog-gravity laboratory, we would be simulating an actual ultracompact object with its corresponding background geometry that exhibits the novel features of an inner horizon or an inner light ring. It does not suffice to engineer a structure that shares only certain isolated features with gravitational spacetimes like an ergoregion or an outer horizon. The main objective of this work is to analyze whether, how, and under which conditions it could be possible to realize the background spacetime \eqref{eq:ST} of a static, spherically symmetric UCO in its entirety or at least part-wise. The structure of the paper is as follows: after this introduction, in Sec. \ref{sec:UCO_ST} we set up the basis for the analogy by writing the sonic metric and drawing a correspondence between fluid and gravitational quantities. In Sec. \ref{sec:4}, we write down the fluid equations and show how the central and the external region of an UCO spacetime can be simulated using specific radial profiles of the external potential and the interaction strength of a Bose-Einstein condensate. Using realistic values from existing BEC experiments, in Sec. \ref{sec:exp} we assess the realizability of our program and list a set of experimental conditions that ensure it. We propose a couple of designs that fulfill all conditions and estimate the timescale on which the instabilities are expected to act for those designs. We conclude with a discussion of the work and an outlook in Sec. \ref{sec:concl}.

\section{\label{sec:UCO_ST}Sonic geometry for RBH/BHM spacetimes}

In this Section, we first present the target gravitational geometry that is to be emulated through an effective acoustic metric that is elaborated upon next. A mapping is established between the two metrics based on a correspondence between condensate and gravitational quantities.

\subsection{The target gravitational metric}

Our target metric \eqref{eq:ST} stems from the fact that regular static, spherically symmetric objects are often presented as a generalization of the Schwarzschild metric, where the ADM mass $M$ is promoted to a one-parameter family of radial mass functions $m(r, \ell)$. The dependence on the regularization parameter $\ell$ will be henceforth suppressed. Popular models are
\begin{equation}
\label{eq:models}
    m(r) =
    \begin{cases}
    \frac{Mr^3}{r^3 + 2M\ell^2} & {\rm Hayward,}\\
        \frac{Mr^3}{(r^2 + \ell^2)^{3/2}} & {\rm Bardeen} \\
        M(1-e^{-r^3/2M\ell^2}) & {\rm Dymnikova.}
    \end{cases}
\end{equation}
A more complete overview of models is given in Table 2 of \cite{mazza2023heart}. In the following, we will frequently use the Hayward model as a working example.

All of these models fulfill one key universal requirement: the mass function is (at least) cubic in radius upon approach of the center $m(r)\to const \cdot r^3$ as $r\to0$, which is the necessary and sufficient condition to obtain finite curvature scalars and thus a regular center. In the following, the Hayward model will be used as a standard example because of its simplicity, but the key points will be valid for all of the above models.

The horizon condition $1-\frac{2m(r)}{r} = 0$ becomes a cubic equation that has two real positive roots (the outer and the inner horizon) for $0 \leq \ell \leq \ell_*$, where e.g. $\ell_* = \frac{4M}{\sqrt{27}}$ for the Hayward model, and no real positive roots for $\ell > \ell_*$, making it a RBH in the former case or a compact object in the latter \cite{mazza2023heart}.

The latter horizonless compact object is a BHM with two light rings until $\ell<\ell_{deg}$, where for Hayward $\ell_{deg}^{H} \approx 0.95M$, as can be inferred from Fig. 2 of \cite{carballorubio2022connectionregularblackholes}. This is the value for which the two light rings become degenerate, after which they disappear. For the Bardeen model there is an analytic result: $\ell_{deg}^{B} = \frac{48}{25\sqrt{5}}M$ \cite{Franzin_2024}. In summary, we have an horizonless UCO that mimics a black hole for $\ell_*<\ell<\ell_{deg}$, which numerically corresponds to
\begin{align}
\label{eq:ltheorval}
\text{Hayward:}\quad & 0.77 \lesssim \ell/M \lesssim 0.95, \\
\text{Bardeen:}\quad & 0.77 \lesssim \ell/M \lesssim 0.86.
\end{align}

Regardless of the horizon structure or lack thereof of the spacetime, it is regular thanks to the presence of $\ell$, which ensures finiteness of curvature scalars at the center $r\to0$. In particular, the Hayward (as well as the Dymnikova) model in Eq. \eqref{eq:models} has the limit $2m(r)/r \to r^2/\ell^2$ as $r\to0$, whence one obtains for the Kretschmann scalar

\begin{equation}
    {\cal K} \equiv R_{\mu\nu\alpha\beta}R^{\mu\nu\alpha\beta} \to \frac{24}{\ell^{4}} \quad {\rm as} \,\, r\to 0
\end{equation}
and surrounding this regular center is a maximally symmetric core region of de-Sitter (dS) type with cosmological constant parameter $\Lambda = \frac{3}{\ell^2}$. Of course, it is easy to construct metrics with Anti-de-Sitter and even Minkowski cores as well, see e.g. \cite{Arrechea:2025nlq, Simpson_2019, Panassiti_2026}. We conclude that $\ell$ becomes a control parameter for the structure of the core of the object. Therefore, in an analog-gravity laboratory, it is enough to identify a tunable experimental parameter or combination thereof as an analogue of $\ell$ to be able to tune the analog UCO's core region. Consequently, the position and instability timescale of features of the internal spacetime region like the inner light ring or the inner horizon come under experimental control. Similarly, in the limit $r\to \infty$, where $m(r)$ asymptotes to the ADM mass $M$ of the spacetime, the condensate quantities serving as analogue for $M$ can be tuned to control the total mass-energy of the target spacetime.

\subsection{Acoustic geometry mapping}
\label{sec:sonicmet}

Analog-gravity programs based on fluids rely on the analogy between the wave equation obeyed by the linear perturbations of the velocity potential $\phi$ of an irrotational velocity flow\footnote{For the extension to flows with vorticity, see \cite{PerezBergliaffa:2001nd}.} $\mathbf{v}=\mathbf{\nabla}\phi$ and the Klein-Gordon equation describing wave excitations of a scalar field on a curved spacetime. Indeed, the wave equation of the former can be written in Klein-Gordon form with an effective geometry given by the so-called \emph{sonic} or also \emph{acoustic} metric with line element \cite{Schuetzhold_Unruh_2002, LivingReviews}
\begin{equation}
\label{eq:sonicmetric}
ds^2_{\textsc{ac}}=-\big(c_s^{2}-v_{r}^{2}\big)\,dt^{2}- 2v_{r}\,dr\,dt + dr^{2}+r^{2}d\phi^{2}\,.
\end{equation}
where $c_s(r)$ is the sound speed and $v_r(r)$ is the radial velocity profile of the condensate.

In this work, we will choose as an experimental reference point BEC experiments where the ultracold atoms are effectively confined to a two-dimensional disc \cite{Viermann_2022, duchene2026maxwellfisheyelensboseeinstein, Tolosa_Sime_n_2022}\footnote{An alternative, whose exploration we leave for future work, are ring-shaped experiments, where the atoms are confined to a 1-dimensional spatial configuration \cite{Garay_2000, Garay_2001, Yatsuta_2020, Jain_2007}.}. For this reason, we have already written the metric in a 2+1 dimensionally reduced form with polar symmetry and purely radial flow $\mathbf{v} = v_r \mathbf{e}_r$. A conformal factor, irrelevant for null \emph{rays} --- which control the light-ring structure and, in the eikonal limit, the long-lived modes that carry the signatures of our instabilities of interest --- has been omitted. We stress that this omission is innocuous at the level of the causal structure and of the geodesic analysis performed below, but not at the level of wave amplitudes: the massless wave equation is conformally invariant only in $1+1$ dimensions, so a quantitative prediction of echo amplitudes or of the full time-domain waveform must retain the conformal factor $\rho^2/c_s^2$ for the 2+1-dimensional case \cite{LivingReviews}.

In order to establish a direct map between the sonic~\eqref{eq:sonicmetric} and the gravitational line elements~\eqref{eq:ST}, it is convenient to rewrite the latter in ingoing Painlev\'e--Gullstrand (PG) form by means of the time redefinition $dt_{\textsc{pg}} = dt + \frac{\sqrt{2m(r)/r}}{1-{2m(r)}/{r}}\,dr$.
under which the metric~\eqref{eq:ST} becomes
\begin{equation}
\label{eq:PGmetric}
ds^2_{\textsc{gr}}
= -\left(1-\frac{2m(r)}{r}\right)dt_{\textsc{pg}}^{2}
+2\sqrt{\frac{2m(r)}{r}}\,dr\,dt_{\textsc{pg}}
+dr^{2}+r^{2}d\Omega^{2}\,.
\end{equation}
We see, as mentioned above, that $g_{rr} = 1$ in these coordinates, while a non-trivial $g_{tr}$ has appeared. Since this metric is in spherical coordinates, to draw the connection to the 2+1-dimensional acoustic metric in polar coordinates, we must simply set $\theta = \pi/2$, placing ourselves on the equatorial plane of the spherically symmetric gravitational metric, effectively yielding $d\Omega^2 \to d\phi^2$. Due to the spherical symmetry of the target geometry, realizing only its equatorial plane is sufficient.

Going back to the sonic metric, we note it is still not in the form necessary to establish an exact correspondence with (\ref{eq:PGmetric}). Since both $c_{s}(r)$ and $v_r(r)$ are in general functions of $r$, one could contemplate the transformation $dt\to dt_{\textsc{ac}} \equiv c_{s}(r)dt$ (an effective local desynchronization of laboratory clocks) in the sonic metric, in order to bring the 2+1-dimensional acoustic line element into the form
    \begin{equation}
    \label{eq:metrtac}
        ds^2_{\textsc{ac}} = -\left(1-\frac{v_r^2}{c_{s}^2}\right)dt_{\textsc{ac}}^2 - 2\frac{v_r}{c_{s}}drdt_{\textsc{ac}} + dr^2 + r^2d\phi^2 \, ,
    \end{equation}
which would allow for a direct comparison with Eq. \eqref{eq:PGmetric}.

It is however important to be precise about the status of this step\footnote{Note that this transformation was also used in a previous work \cite{Pomakov:2026alh} proposing a similar program for surface-gravity waves on fluids. There, we focused entirely on the background realization and did not delve into the precise light ring structure, which is where this step comes with a caveat, as we shall see.}. In fact, the one-form $c_s(r)\,dt$ is not closed, $d\left(c_s dt\right)=c_s'\,dr\wedge dt\neq0$, so for a non-constant $c_s(r)$ there exists no function $t_{\textsc{ac}}$ whose differential it is. This means that $t_{\textsc{ac}}$ \emph{cannot be considered a valid time coordinate} but one has to rather view Eq.~\eqref{eq:metrtac} as the acoustic metric \eqref{eq:sonicmetric} written in an anholonomic coframe.\footnote{A coframe $\{\theta^a = e^a{}_\mu\, dx^\mu\}$ is called \emph{anholonomic} if it cannot be written as a set of coordinate differentials, i.e.\ there are no functions $y^a$ such that $\theta^a = dy^a$.} We will analyze the precise effects of this in the next subsection.

Let us make the following identification between condensate and gravitational quantities by comparing the corresponding line elements (\ref{eq:metrtac}) and (\ref{eq:PGmetric}):

\begin{equation}
\label{eq:identif}
    \frac{v_r(r)}{c_s(r)} = - \sqrt{\frac{2m(r)}{r}}
\end{equation}
Clearly, to emulate an UCO in a condensate, we need a negative radial velocity, i.e., an inflow towards the center. In contrast, an outflow would mimic white-hole-like behavior.

\subsection{The residual sound-speed factor and the analog light ring structure}
\label{sec:residual}

Imposing only the correspondence \eqref{eq:identif}, it is illustrative to bring back the acoustic line element \eqref{eq:sonicmetric} to diagonal form, by the \emph{bona fide} coordinate change $d\tau = dt + \frac{v_r}{c_s^2-v_r^2}dr$ (whose coefficient is a function of $r$ alone, hence integrable). One finds
\begin{equation}
\label{eq:acdiag}
ds^2_{\textsc{ac}} = -\,c_s^2(r)\,f(r)\,d\tau^2 + \frac{dr^2}{f(r)} + r^2 d\phi^2\,,
\qquad f(r)\equiv 1-\frac{v_r^2}{c_s^2}=1-\frac{2m(r)}{r}\,,
\end{equation}
to be compared with the equatorial section of the target \eqref{eq:ST}, $-f\,dt^2+f^{-1}dr^2+r^2d\phi^2$. The two agree in $g_{rr}$ and $g_{\phi\phi}$ and differ only by the factor $c_s^2(r)$ multiplying $g_{\tau\tau}$. Since $g_{\phi\phi}$ fixes the areal radius and $g_{rr}$ is already in the target form, this factor can be removed neither by a further coordinate change nor by a conformal rescaling, unless $c_s$ is constant. Its effects must therefore be assessed feature by feature. Let us list them below.
\begin{itemize}
  \item \emph{Horizons are preserved:} The condition $f=0$ is $|v_r|=c_s$: the position of $r_\pm$, and hence the entire causal structure (trapped region, inner and outer horizon, regular core) is reproduced without any correction.
  \item \emph{Surface gravities are exact up to a local unit conversion.} For a metric $-A\,d\tau^2 + B\,dr^2$ one has $\kappa = \frac{1}{2}A'/\sqrt{AB}$, so that with $A=c_s^2f$, $B=1/f$, the acoustic surface gravity becomes $\kappa_{\textsc{ac}, \pm} = c_s(r_\pm)\,\kappa_{\textsc{gr}, \pm}$. The extra sound speed factor is exactly the one needed to convert the gravitational surface gravity $\kappa_{\textsc{gr}}$ (an inverse length) into an inverse laboratory time; it is accounted for in Eq.~\eqref{eq:tauih} below. Note that $\kappa_{\textsc{gr}}$ does actually include the analogous light speed factor $c$, but it is implicit since we are working in units $c=1$.
  \item \emph{Light rings are displaced:} For a stationary metric $-A\,d\tau^2+B\,dr^2+r^2d\phi^2$, circular null orbits satisfy $\left( A/r^2\right)'=0$, independently of $B$. In the target, where $A = f$, this is the familiar $f'/f = 2/r$. In the analog geometry \eqref{eq:acdiag}, where instead $A = c_s^2 f$, the same computation gives
  \begin{equation}
  \label{eq:LRanalog}
  \frac{f'(r)}{f(r)} \;=\; \frac{2}{r}\Big[1-\sigma(r)\Big]\,,
  \qquad
  \sigma(r)\;\equiv\;\frac{d\ln c_s}{d\ln r}\,,
  \end{equation}
  where $\sigma$ is the local logarithmic slope of the sound-speed profile. Note that nothing in this statement refers to a specific analog platform: in any acoustic spacetime built on a spatially varying propagation speed the light rings are displaced with respect to those of the gravitational target, by an amount governed by $\sigma$ alone.
\end{itemize}
This equation has, in general, multiple solutions, that are collectively denoted $r_{\textsc{lr}}$. In the case of a BHM of the Hayward, Dymnikova or Bardeen type, it has two solutions: $r_{\textsc{ilr}}$ and $r_{\textsc{olr}}$, i.e., the position of the inner and the outer light ring, respectively.

We now see that, to ensure the light rings sit at their gravitational positions, it is necessary and sufficient to make $\sigma(r_{\textsc{lr}})=0$. Instead, an even stronger condition would be $\sigma(r) = 0$ for all $r$, indicating a spatially uniform propagation speed, which would erase any difference in lapse between the gravitational metric \eqref{eq:ST} and the acoustic metric \eqref{eq:acdiag}, not just ensure coinciding light ring positions. Put differently, if $\sigma(r) = 0$ can be satisfied globally, the now constant $c_s^2$ factor \emph{can} be absorbed by a proper coordinate transformation $dt \to dt_{\textsc{ac}} = c_sdt$ and the analogy between the metrics Eqs. \eqref{eq:metrtac} and  \eqref{eq:PGmetric} becomes exact, provided the identification Eq. \eqref{eq:identif}.

Let us also note that displacement is modest but not negligible. To gauge its size, take a Schwarzschild target $m(r)=M$, which has a single (outer) light ring, and a sound speed behaving locally as a power law, $c_s\propto r^{\gamma}$ with constant $\gamma$. Equation~\eqref{eq:LRanalog} is then solved in closed form,
\begin{equation}
\label{eq:LRschw}
r_{\textsc{olr}} \;=\; \frac{3-2\gamma}{1-\gamma}\,M\,,
\end{equation}
which returns the familiar $3M$ at $\gamma=0$, moves the light ring inwards when the sound speed decreases outwards ($\gamma<0$) and outwards when it grows ($\gamma>0$).

Whether $\sigma$ can be made to vanish (be it only at $r_{\textsc{lr}}$ or globally), and how large the displacement is when it cannot, are questions about the specific laboratory system: both are settled in Sec.~\ref{sec:LRcond}, once the sound-speed profile of the condensate has been derived.

\subsection{Simulating the horizon structure and the core region via the condensate flow}

\begin{figure}
\includegraphics[width=10cm]{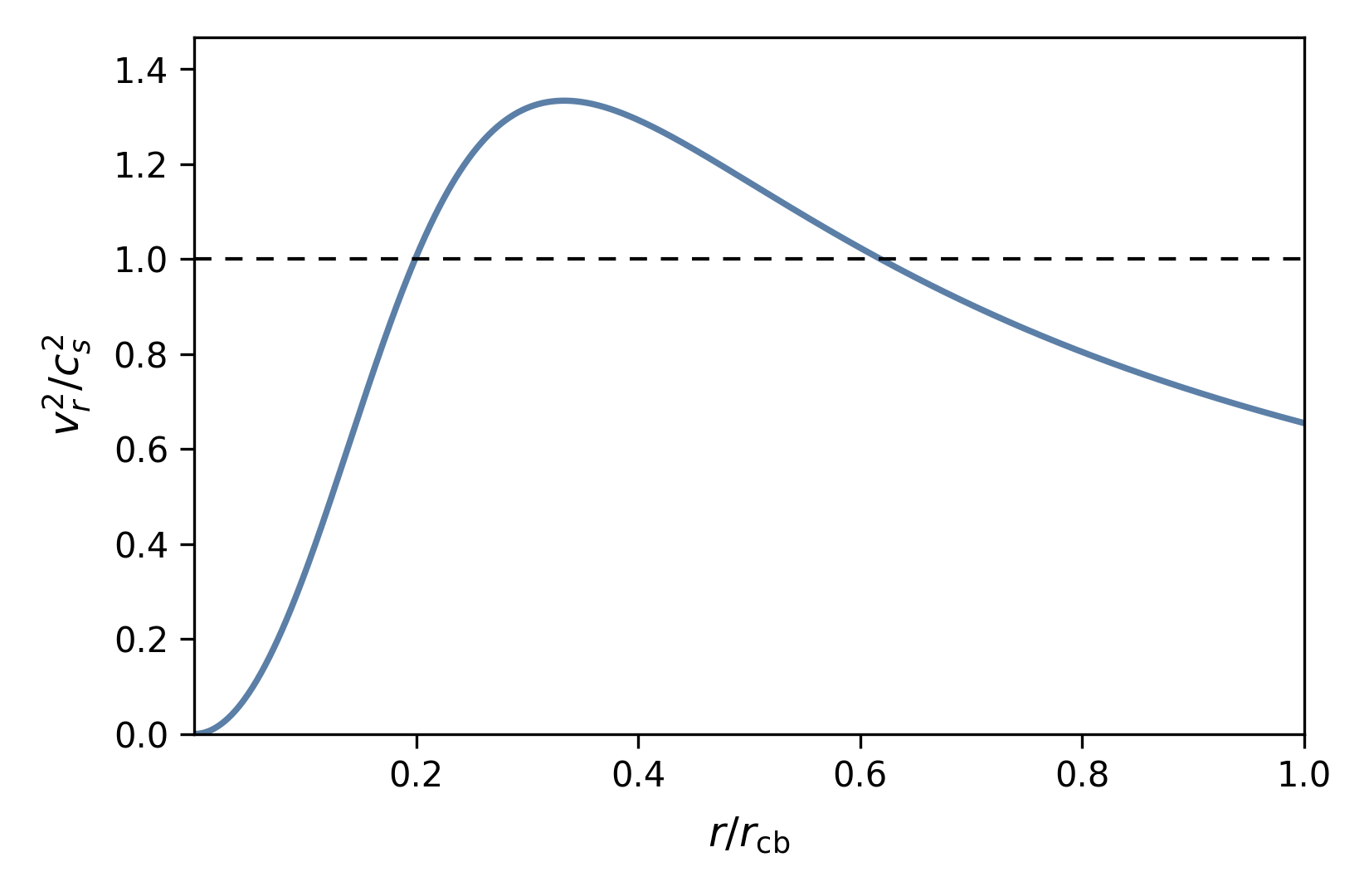}
\caption{The desired behavior of $(v_r/c_s)^2$ versus radius $r$, normalized to the radius $r_{\rm cb}$ of the container wall, to realize an analog Hayward UCO.}
\label{fig:vcssq}
\end{figure}

To show how through a given radial velocity and sound speed profile one simulates the horizon structure of an UCO spacetime, consider the plot in Fig. \ref{fig:vcssq}, where the squared ratio $v_r(r)/c_{s}(r)$ is shown for an ultracompact object of Hayward type. It exhibits a peak that is larger/smaller than unity for a RBH/BHM. For a peak position lower than unity, there would be no trapped region and the object would be a horizonless BHM. If the peak were precisely at value unity, the object would be an extremal RBH with two degenerate horizons. In this plot, the peak is larger than unity, making it a RBH, with the outer and inner horizon located at the two crossings $r = r_{\pm}$ of unity: the outer and the inner horizon of a RBH. Between the two horizons, the region of spacetime is trapped (i.e. supersonic). Outside of the outer horizon the region is obviously untrapped (subsonic), and, more interestingly, it is untrapped (subsonic) also in the interior of the inner horizon, which is the regular de-Sitter core. In the BHM case, of course, there is no trapping and therefore no regions of supersonic flow.

For a given UCO model of choice, $m(r)$ is a known function of radius. To realize this particular model, a certain $r$-dependence of $v_r(r)$ and/or $c_s(r)$ must be realized, so that their ratio gives $(- \sqrt{2m(r)/r})$; this is the gist of Eq. (\ref{eq:identif}). In a BEC with density profile $\rho(r)$ and particle mass $m_p$, $c_s(r)=\sqrt{g(r)\rho(r)/m_p^2}$ is under direct experimental control thanks to the control over the interaction strength $g(r)$ by means of an external magnetic field through the Feshbach resonance \cite{LivingReviews, Balbinot_2008, Steinhauer_2016, Macher_2009, liberati2009analoguemodelsemergentgravity}. This magnetic field can be made to vary with space in the desired manner, so that the identification (\ref{eq:identif}) is fulfilled \cite{LivingReviews, Sierant_2017, bagnato2015boseeinsteincondensationyears}.

Since both the inner horizon (in the RBH case) and the stable inner light ring (for a BHM) are typically located near the boundary of the maximally symmetric core, we will mostly be interested in the realization of the core region. For all models of $m(r)$ under consideration
\begin{equation}
\label{eq:behavatcenter}
    \frac{v_r(r)}{c_{s}(r)} = - \sqrt{\frac{2m(r)}{r}} \to -\frac{r}{\ell} \propto r, \quad {\rm as} \,\,r \to 0
\end{equation}
We conclude that the analog regularization parameter can be tuned via ratio between fluid velocity and sound speed. This ratio, in turn, will be controlled through the external potential and the interaction strength of the ultracold atoms, as we will show in the next section.

The asymptotic spacetime region $r\to \infty$ can be probed close to the condensate boundary $r=r_{\rm cb}$ if $r_{\rm cb} \gg M$. In this case, we would need the behavior

\begin{equation}
\label{eq:behavatinfty}
    \frac{v_r(r)}{c_{s}(r)} \to -\sqrt{\frac{2M}{r}} \propto r^{-1/2}, \quad {\rm as} \,\,r \to \infty
\end{equation}
This is also a model-independent statement, in that for all choices of $m(r)$ it must asymptote to $M$ as an expression of the requirement of asymptotic flatness. In conclusion, the analog of $M$ will be found in the constant factor in front of a $r^{-1/2}$ fall-off of the ratio between radial flow velocity and sonic speed at the external parts of the experiment.

Let us note that apart from emulating the two limiting behaviors of an UCO, i.e.~$m(r)\to \frac{r^3}{2\ell^2}$ towards the center, and $m(r) \to M$ at infinity, with the methods we illustrate here one could also simply simulate a Schwarzschild black hole. To do this, one must simply realize the asymptotic regime $m(r)\to M$ globally.

The goal of the next section will be to solve the continuity and Bernoulli equations along with the correspondence equation (\ref{eq:identif}) in the interior and the asymptotic spacetime regions, with the goal of determining the external potential and interaction strength required to emulate these parts of the geometry. We will identify the analogues of the gravitational parameters $\ell$ and $M$ among the constant factors within the external potential.

\section{Analog RBH/BHM in a BEC}
\label{sec:4}

As was mentioned, the test-field wave perturbations on the BEC behave as if propagating on a polar-symmetric 2+1-dimensional curved spacetime with metric \eqref{eq:metrtac}, since the system in the laboratory is kept spatially confined to a disc \cite{duchene2026maxwellfisheyelensboseeinstein, Viermann_2022, Tolosa_Sime_n_2022}. The independent variable in this disc-symmetric problem is thus the radial position $r$. The system is considered to be in the fluid regime, governed by the continuity and the Bernoulli equation \cite{LivingReviews}. We furthermore want the correspondence equation (\ref{eq:identif}) to be fulfilled. 

The variables in these equations are radial velocity of the condensate $v_r(r)$, the sound speed $c_s(r) = \sqrt{g(r) \rho(r)/m_p^2}$ or equivalently the interaction strength $g(r)$, the mass density $\rho(r)$, and an external potential $V_{\rm ext}(r)$ that is used to control the system. The total number of dependent variables entering the hydrodynamical equations is thus four: $(v_r(r), \rho(r), V_{\rm ext}(r), g(r))$, and there are three equations: the continuity, Bernoulli and the correspondence equation.

Two of the variables are under direct experimental control: $V_{\rm ext}(r)$ and $g(r)$. It could be that it is unnecessary to control $g(r)$ by means of a spatially varying magnetic field and the Feshbach resonance, and that the three variables $(v_r(r), \rho(r), V_{\rm ext}(r))$ are enough to satisfy three equations. In essence, one would choose $V_{\rm ext}(r)$ such that, through the Bernoulli and the continuity equation, $v_r(r)$ and $\rho(r)$ take on the form required to satisfy the correspondence equation, solving the set of three equations consistently with the three variables.

Nevertheless, the freedom to vary $g(r)$ as a fourth control variable provides a useful tool to ensure the analog shares the same light ring structure as the gravitational target --- we shall derive in Sec.~\ref{sec:LRcond} the profile that achieves this --- although we already discussed this is not a necessity. Similarly, it can be used (see Sec.~\ref{subsec:coresol}) to boost the contribution of enthalpy with respect to the quantum pressure term, which will be shown to erase sensitivity of the experiment to the analog $\ell$ parameter. However, also here the variation of $g(r)$ is not strictly necessary because enthalpy exceeds the quantum pressure whenever $r$ is a few times the local healing length, so a spatially constant $g$ already retains an analogue of $\ell$ throughout the region that an experiment can resolve. Indeed, Sec.~\ref{sec:exp} will show that a \emph{strongly} varying $g$ is excluded by the requirements of quasi-two-dimensionality and of a sufficiently long three-body lifetime. The freedom of a slowly varying $g(r)$ for extra experimental control thus remains, but is not forced upon us.

Let us close this section by noticing that the above discussed simple "variable-counting" argument illustrates the considerably greater potential of BEC analog systems for our purposes with respect to the surface-gravity fluid systems we studied previously in \cite{Pomakov:2026alh}. There, we had fewer variables than equations, which made it possible to satisfy them all at the same time essentially "by chance" and under several approximations.

\subsection{Continuity, Bernoulli and correspondence equation}
\label{subsec: fluideqs}

We now lay out the equations governing the system. We begin with the continuity equation, which in a stationary flow, where partial time derivatives of any quantity vanish $\partial_t = 0$, and in cylindrical symmetry with $v_z = v_{\phi}=0$, takes on the form

\begin{equation}
\label{eq:cont}
    \partial_r(\rho v_r r)=\frac{d}{dr}(\rho v_r r) \equiv (\rho v_r r)'=0 \, \,.
\end{equation}
At the same time, we want the radial velocity to satisfy the correspondence equation which we restate
\begin{equation}
\label{eq:correq}
    v_r(r) = - \sqrt{\frac{2m(r)}{r}}c_s(r)
\end{equation}
Finally, we have the Bernoulli equation per particle mass for a stationary flow
\begin{equation}
\label{eq:bern}
    \frac{1}{2}v_r(r)^2 + \frac{V_{\rm ext}(r)}{m_p} + h(r) + Q(r) = \frac{\mu}{m_p}
\end{equation}
where $h(r)$ is the enthalpy per mass, $Q(r)$ is the quantum pressure term, and the total constant energy-per-mass on the r.h.s. is called $\frac{\mu}{m_p}$ since for a BEC it corresponds to the chemical potential per mass.

If we now introduce the short-hand notation $\frac{V_{\rm ext}(r)}{m_p} \equiv U(r)$ for the external potential per mass and pack all terms apart from the kinetic energy density into an \emph{effective potential}\footnote{The constant term $ \frac{\mu}{m_p}$ can be absorbed into the external potential $U(r)$ as a constant shift.}

\begin{equation}
    U_{\rm eff}(r) \equiv U(r) + h(r) + Q(r)\,,
\end{equation}
then we can immediately write that, to satisfy the correspondence equation (\ref{eq:correq}), we need the effective potential of the given fluid system to satisfy

\begin{equation}
\label{eq:ueff}
    U_{\rm eff}(r) = - \frac{m(r)}{r}c_s^2(r)
\end{equation}
Strong of these conditions we shall now move on to discuss the features that they implies on our elective experimental system, i.e.~on a disc shaped Bose--Einstein condensate.

\subsection{The Bose--Einstein condensate}

\subsubsection{Physical quantities and features}
We previously listed the fluid equations and the correspondence governing the analogy, which would be valid for any fluid system. In the specific case of a BEC confined to a disc, the following particularities apply
\begin{itemize}
    \item The mass density $\rho(r)$ and interaction strength $g(r)$ are now two-dimensional, since the condensate is tightly trapped in the $z$-direction, resulting in an effectively dimensionally-reduced $2D$ spatial geometry. This changes nothing in the formulas used, but for the experimental estimates in Sec. \ref{sec:exp} we need to keep in mind that the dimensions of these quantities are now mass/length$^2$ and mass$\,\times \,$length$^4$/time$^2$.
    \item The sound velocity $c_s^2(r) = \frac{g(r)\rho(r)}{m_p^2}$ is in general not constant in radius and can be controlled to a certain degree via the tuning of the interaction strength $g(r)$ by means of the Feshbach resonance. This complicates the mapping between fluid and gravitational quantities but ensures a lot of experimental control.
    \item The enthalpy per mass is also not constant anymore and can be calculated as follows:
    \begin{equation}
    \label{eq:enthal}
        h(r) = \int_0^{\rho}\frac{d\tilde{\rho}}{\tilde{\rho}}c_s^2 = \frac{g(r)\rho(r)}{m_p^2} = c_s^2(r)  \, .
    \end{equation}
    \item There is a non-zero quantum pressure (per mass) given, in cylindrical coordinates, by the expression
    \begin{equation}
        Q(r) = -\frac{\hbar^2}{2m_p^2}\frac{1}{\sqrt\rho}\frac{1}{r} \frac{\partial}{\partial r}\left( r \frac{\partial\sqrt\rho}{\partial r} \right) \, ,
    \end{equation}
    where $\rho(r)$ must be expressed via the continuity equation, which we do in the following.
\end{itemize}

\subsubsection{General solution}

Inserting the sound speed into the radial velocity \eqref{eq:correq} required to ensure the analogy between gravitational and acoustic spacetimes, we obtain
\begin{equation}
\label{eq:radvel}
    v_r(r) = - \sqrt{\frac{2g(r)m(r)\rho(r)}{m_p^2\,r}}
\end{equation}
Inserting this into the continuity equation (\ref{eq:cont}) yields

\begin{equation}
\label{eq: dens}
    \rho(r) = \frac{B}{(rg(r)m(r))^{1/3}}.
\end{equation}
with an integration constant $B$, related to the flux of inflowing fluid, see Sec. \ref{sec:exp}. Here we see that to get a real density, we need a positive (repulsive) interaction strength $g(r)$. Now we need to insert this into the expressions for the quantum pressure, the effective potential and the enthalpy, to obtain the $U(r)$ to be imposed on the system. Starting with the quantum pressure, one gets
\begin{equation}
\label{eq:qp}
    Q(r) = -\frac{\hbar^2}{2m_p^2} \left( \frac{1}{36r^2} + \frac{7}{36}\frac{m'^2}{m^2} - \frac{1}{9r}\frac{m'}{m}  - \frac{1}{6}\frac{m''}{m} + \frac{1}{18}\frac{m'g'}{mg} + \frac{7}{36}\frac{g'^2}{g^2} - \frac{1}{9r}\frac{g'}{g}  - \frac{1}{6}\frac{g''}{g} \right)
\end{equation}
Now, using $c_s^2=g\rho/m_p^2$ and eliminating the density via eq. (\ref{eq: dens}) in the expression (\ref{eq:ueff}) for the effective potential (equivalently kinetic energy), gives
\begin{equation}
    U_{\rm eff}(r) \equiv -\frac{1}{2}v_r^2(r) = -\frac{B}{m_p^2}\frac{1}{r^{2/3}}\left(\frac{m(r)g(r)}{r}\right)^{2/3}
\end{equation}
Finally, inserting the density expression into the enthalpy given by (\ref{eq:enthal}), one finds
\begin{equation}
\label{eq:enth}
    h(r) = \frac{B}{m_p^2}\frac{g(r)^{2/3}}{m(r)^{1/3}r^{1/3}} = c_s^2(r)
\end{equation}
which was also found to be the sound speed squared.

In the end, for the actual external potential per mass used in the experiment, $U(r) = U_{\rm eff}(r)-h(r)-Q(r) + \frac{\mu}{m_p}$, where we have reinstated the constant shift for completeness, so that one gets
\begin{equation}
\label{eq:ubec}
    U(r)=\frac{\mu}{m_p}-\frac{B}{m_p^2}\frac{1}{r^{2/3}}\left(\frac{m(r)g(r)}{r}\right)^{2/3} \left(1+\frac{r}{m(r)}\right)+\frac{\hbar^2}{2m_p^2} \left( \frac{1}{36r^2} +.... \right)
\end{equation}
where for legibility we have written explicitly only the first term from the quantum pressure. Due to the structure of the omitted derivative terms, if $m(r)$ and $g(r)$ are power laws in $r$, these terms will also go like $1/r^2$. Furthermore, if the exponent of the power law is of $\mathcal{O}(1)$, they will contribute the same order of magnitude as the first term, and thus the entire quantum pressure will be of that same order of magnitude. Therefore, the first quantum-pressure term stands as "representative" of the order of magnitude of the entire $Q(r)$.

We see that the external potential needs to closely balance between these various contributions in order to yield the just right density profile necessary to simulate an UCO spacetime. In full generality this is a tight balance that is difficult to realize. Nonetheless, things simplify in the two regimes of interest, $r\to0$ and $r\to\infty$, since in each case only one of the three contributions (kinetic energy, enthalpy and quantum pressure) will dominate.



\subsubsection{Solution in the core region}\label{subsec:coresol}

If we now take the limit $r\to0$, meaning $m(r) \to \frac{r^3}{2l^2}$, we expect the quantum pressure to be dominant, \emph{if} $g$ in this region is assumed nearly constant\footnote{As we mentioned, the effect of a non-constant $g(r)$ on the quantum pressure is only to modify the numerical coefficient in front of the $r^{-2}$ behavior. What will end up being more important, as we will see shortly, is its effect on the enthalpy term.}, the necessary external potential per mass to simulate a static, spherically symmetric RBH or BHM becomes
\begin{equation}
\begin{aligned}
    U(r)\to\frac{2\hbar^2}{9m_p^2r^2}
\end{aligned}
\end{equation}
The fact that the quantum pressure term dominates erases any sensitivity to the regularization parameter $\ell$: it does not appear anymore in any laboratory quantity. If we have no analogue of $\ell$ in the center of the simulated spacetime, we do not have control over the core of the object we are trying to simulate and of its instabilities.

This makes intuitive sense: the quantum pressure forces small disturbances in the system to heal, smoothening out the condensate wave function on length scales close to the healing length
\begin{equation}
\label{eq:healing}
\xi = \frac{\hbar}{\sqrt{2g\rho}} \, ,
\end{equation}
which is defined precisely as the scale below which the quantum pressure starts dominating. In this sense, in a BEC, this becomes the length scale that regularizes the system (an analogue of a quantum-gravity regularization scale of sorts), and there cannot be a smaller scale $\ell$ that plays this role.

This problem is easily evaded for the following reason. The limit $r \to 0$ is not to be taken literally as going to sub-healing length scales. Rather, for us $r\to0$ will be tantamount to taking $r\ll M$, or, even more precisely
\begin{align}
\label{eq:centrdef}
  r\lesssim \ell \, \, \, {\rm for} \, \, \, \ell\ll M \, \, \, {\rm and} \, \, \, 2M \lesssim r_{\rm cb} \\
  r \ll \ell \, \, \, {\rm for} \, \, \, \ell\sim M \, \, \, {\rm and} \, \, \, 2M \lesssim r_{\rm cb}
\end{align}
where $r_{\rm cb}$ is the container or confinement boundary (more precisely, the size of the region to which the cold atoms are confined). In both cases, $2M$ must be sufficiently smaller than $r_{\rm cb}$ to see within the experiment a sufficiently large region of the target spacetime. Then, if $\ell$ is sufficiently bigger than the healing length $\xi$ and the system is probed at $r$ sufficiently larger than $\xi$, we will never be in a regime of quantum-pressure dominance with respect to enthalpy.

We can make the above statements more quantitative with the following simple argument. Using $h=c_s^2$ from Eq.~\eqref{eq:enthal} and the first term of the quantum pressure \eqref{eq:qp} taken as representative of its order of magnitude, and recalling $\xi = \hbar/\sqrt{2g\rho}=\hbar/(\sqrt{2}m_pc_s)$, we obtain the ratio
\begin{equation}
\label{eq:hoverQ}
\frac{h(r)}{|Q(r)|}\;\simeq\;36\left(\frac{r}{\xi(r)}\right)^{2}\,,
\end{equation}
so that the quantum pressure takes over only at radii comparable to the healing length itself. Evaluating the full expression \eqref{eq:qp}, rather than only its first term, reduces the coefficient by about an order of magnitude in the core, where the derivative terms in $m(r)$ contribute. For the fiducial designs of Sec.~\ref{sec:exp} one finds $h/|Q| \gtrsim 10$ at every radius down to the smallest one at which the condensate is still quasi-two-dimensional. In other words, the regime of quantum-pressure dominance lies below the range of validity of the whole description and is never reached in practice. This is the precise sense in which the problem can be evaded.

With enthalpy dominating over quantum pressure (and chemical potential per mass), in the limit $r\to0$ to be understood as explained above, the required $U(r)$ takes on the form 
\begin{equation}
\label{eq:potcoregeneral}
    U(r)\sim - \frac{2^{1/3}Bg^{2/3}(r)l^{2/3}}{m_p^2r^{4/3}} \propto -r^{-4/3}g^{2/3}(r)
\end{equation}
We see that we retain sensitivity to and control of $\ell$, it being a constant parameter within the external potential that can be tuned.

It is particularly easy to experimentally realize $g(r) \propto r^{-2}$ via a dipolar magnetic coil trap, so let us take this as a numerical example, postponing the discussion of whether this behavior is viable for our programme to Sec. \ref{sec:exp}. In this case the required potential will go like
\begin{equation}
\label{eq:potcoreprop}
U(r) \propto -r^{-8/3} \,\,
\end{equation}
from which the result for the density and velocity profiles follow from the fluid equations as
\begin{align}
    \rho(r) \sim \frac{2^{1/3}Bl^{2/3}}{g^{1/3}(r)r^{4/3}} \propto r^{-4/3}g^{-1/3}(r) \propto r^{-2/3} \\  v_r(r) \sim -\frac{2^{1/6}B^{1/2}r^{1/3}g^{1/3}(r)}{l^{2/3}m_p} \propto r^{1/3}g^{1/3}(r) \propto r^{-1/3} \\
    c_s(r) \sim \frac{2^{1/6}B^{1/2}g^{1/3}(r)l^{1/3}}{m_pr^{2/3}} \propto r^{-2/3}g^{1/3}(r) \propto r^{-4/3}
\end{align}

One might object that to simulate a regular gravitational object we are relying on divergent quantities in the analog platform, which seems counter-intuitive, let alone experimentally unfeasible. This has again to do with how literally one interprets the $r\to 0$ limit. We are not interested in the exact center $r=0$ and it will be necessarily regular in any analog-gravity platform. In some analog-gravity experiments, for example, there is a sink in the center to conduct a radially inflowing fluid away, in which case it must be refilled into the container elsewhere \cite{Smaniotto_2025, Svancara_2024}. Delegating these experimental considerations to Sec. \ref{sec:exp}, we briefly restate that we are interested in the region $r\lesssim l$ around the center, whereas in the exact center, the divergences in the quantities will certainly be smoothed out by the experimental setup.

\subsubsection{Solution in the external region}

Moving on to the asymptotic regime, $r\to \infty$\footnote{We recall that this is short-hand for  $r\to r_{\rm cb}$ for $r_{\rm cb} \gg M$}, where $m(r) \to M$, we see from eq. (\ref{eq:ubec}) that the enthalpy term will fall off to zero the slowest and thus again be the dominant term, yielding the potential to be imposed there up to the constant shift
\begin{equation}
\label{eq:potasymp}
    \left[U(r\to \infty) - \frac{\mu}{m_p}\right]\to -\frac{Bg^{2/3}(r)}{m_p^2M^{1/3}r^{1/3}} \propto -r^{-1/3}g^{2/3}(r)
\end{equation}

Once more, the gravitational spacetime's ADM mass $M$ can be extracted from the constant factor in the external potential. Inserting $g(r) \propto r^{-2}$ merely as a numerical example like in the core region we obtain
\begin{equation}
\label{eq:potext}
    U\propto -r^{-5/3} \, .
\end{equation}

The corresponding density and velocity profiles from the fluid equations are
\begin{align}
    \rho(r) \sim \frac{B}{M^{1/3}g^{1/3}(r)r^{1/3}} \propto r^{-1/3}g^{-1/3}(r) \propto r^{1/3} \\  v_r(r) \sim -\frac{2^{1/2}B^{1/2}M^{1/3}g^{1/3}(r)}{r^{2/3}m_p} \propto r^{-2/3}g^{1/3}(r) \propto r^{-4/3} \\
    c_s(r) \sim \frac{B^{1/2}g^{1/3}(r)}{m_pM^{1/6}r^{1/6}} \propto r^{-1/6}g^{1/3}(r) \propto r^{-5/6}
\end{align}
If we realize this behavior of the external potential and the interaction strength globally and not just at $r\gg \ell$, we would obtain an analog Schwarzschild black hole.

\subsubsection{The analog light-ring structure and how to restore it}
\label{sec:LRcond}

We are now in a position to return to the light-ring displacement identified in Sec.~\ref{sec:residual}, which was left there as a statement about a generic sound-speed profile. In the condensate that profile is no longer free: the continuity equation together with the correspondence Eq.\eqref{eq:identif} fixes $c_s^2 = \frac{B}{m_p^2}\,g^{2/3}(r\,m)^{-1/3}$, cf. Eq.~\eqref{eq: dens}, so that the logarithmic slope entering Eq.~\eqref{eq:LRanalog} is controlled by the interaction strength and the mass function alone,
\begin{equation}
\label{eq:sigmaBEC}
\sigma(r) \;=\; \frac{1}{3}\frac{d\ln g}{d\ln r} \;-\; \frac{1}{6}\left(1+\frac{d\ln m}{d\ln r}\right)\,.
\end{equation}
We can now quote the numbers for the cases of interest. Specializing as above to $g\propto r^{-a}$, for which the asymptotic slope is $\sigma\to-(1+2a)/6$ and Eq.~\eqref{eq:LRschw} becomes $r_{\textsc{olr}}=4M(5+a)/(7+2a)$: this amounts to $r_{\textsc{olr}}=20M/7\simeq2.86M$ for $a=0$ and $28M/11\simeq2.55M$ for $a=2$, against the standard Schwarzschild gravitational result $3M$. 

\begin{figure}
\includegraphics[width=14.5cm]{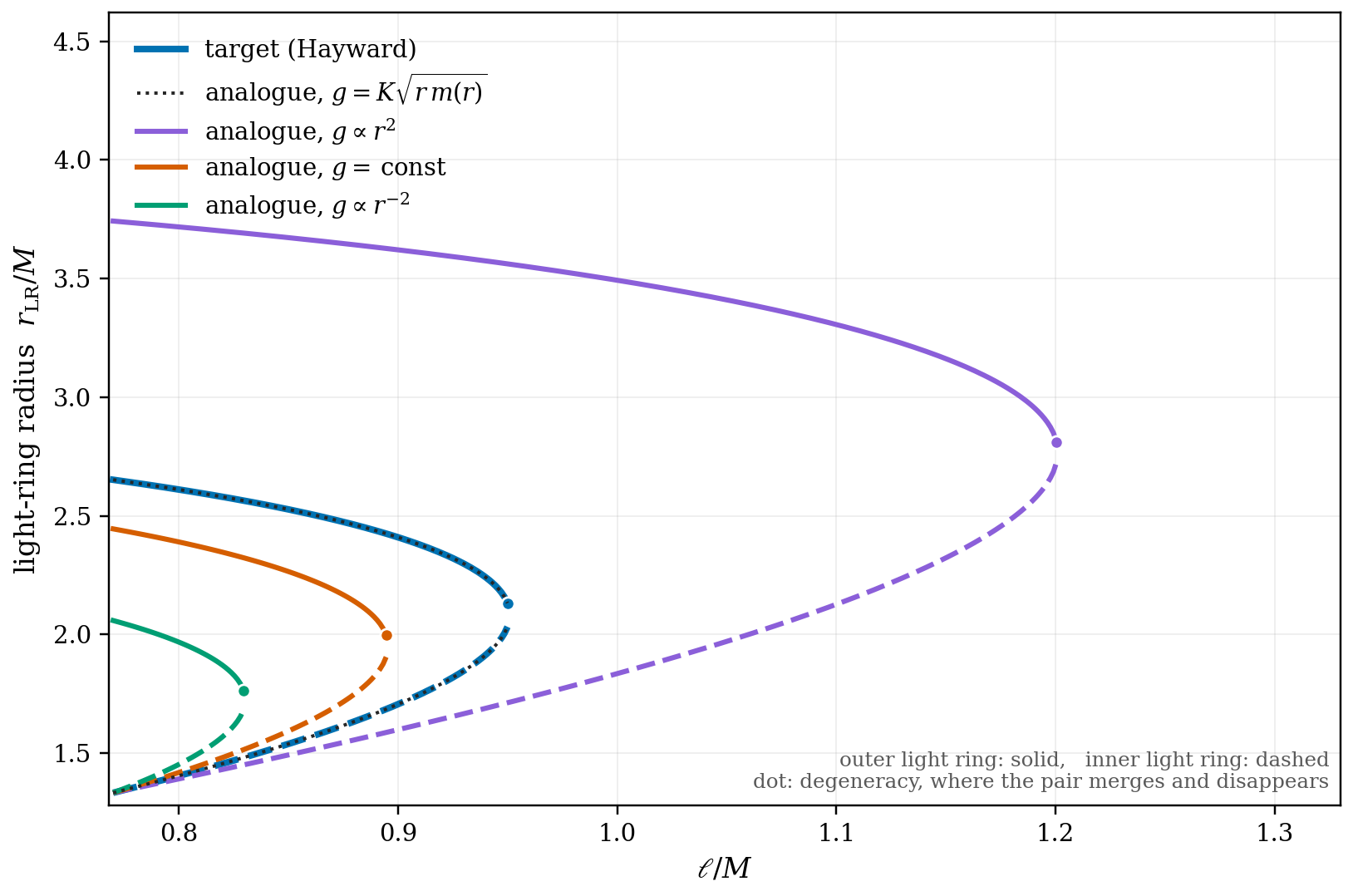}
\caption{Light-ring radii of the Hayward one-parameter family as a function of $\ell/M$: the gravitational target (blue), and the analog geometry \eqref{eq:acdiag} for $g\propto r^{2}$ (purple), for a constant interaction strength (orange) and for $g\propto r^{-2}$ (green). Solid: outer light ring; dashed: inner light ring; dots mark the degeneracy at which the pair merges and disappears. The residual factor $c_s^2(r)$ leaves the horizon threshold $\ell_*$ untouched but moves the upper edge of the two-light-ring window, Eq.~\eqref{eq:lwindow}, downwards for a sound speed that decreases outwards and upwards for one that grows. The black dotted curve is the prescription \eqref{eq:gslnretainLRgrav}, $g=K\sqrt{r\,m(r)}$: it has $\sigma\equiv0$ and therefore lies on the gravitational target identically, by construction.}
\label{fig:lrwindow}
\end{figure}

In the BHM case, more importantly, the displacement is not the same for the two light rings, so it changes the value of $\ell$ at which they merge and disappear. Figure~\ref{fig:lrwindow} shows the two branches as functions of $\ell/M$ for the Hayward model. The lower edge of the window is set by the horizon condition and is therefore untouched, $\ell_*/M = 4/\sqrt{27}\simeq0.770$, whereas the upper edge moves, downwards for $\sigma<0$ and upwards for $\sigma>0$:
\begin{align}
\label{eq:lwindow}
\text{target (Hayward):}\quad & 0.77 \lesssim \ell/M \lesssim 0.95\,, \nonumber\\
\text{analog},\;g\propto r^{2}:\quad & 0.77\lesssim \ell/M \lesssim 1.20\,,\nonumber\\
\text{analog},\;g={\rm const}:\quad & 0.77 \lesssim \ell/M \lesssim 0.90\,, \\
\text{analog},\;g\propto r^{-2}:\quad & 0.77\lesssim \ell/M \lesssim 0.83\,.\nonumber
\end{align}
These shifts have a very relevant implication: an experiment tuned to $\ell/M=0.92$ with a uniform interaction strength, which in general relativity is a perfectly good mimicker with two light rings, would produce an analog object with \emph{no} light rings at all; conversely, at $\ell/M=1.1$ with $g\propto r^2$ the analog system would possess a light-ring pair, and hence an echo signal, that the gravitational target it is supposed to emulate does not have (see again Figure~\ref{fig:lrwindow}). In what follows, whenever we quote light-ring positions or echo delays for the analog system we use Eq.~\eqref{eq:LRanalog} rather than the gravitational condition.

We can now make the criterion $\sigma\equiv0$ concrete by making use of the fact that $g(r)$ is a free function under experimental control and can then be chosen so to preserve the gravitational light-ring structure $f'/f=2/r$ in our analogue set-up. By Eq.~\eqref{eq:sigmaBEC} this amounts to requiring
\begin{equation}
\label{eq:retainLRgrav}
 \frac{d\ln g}{d\ln r} \;=\; \frac{1}{2}\left(1+\frac{d\ln m}{d\ln r}\right) \, ,
\end{equation}
which integrates to
\begin{equation}
\label{eq:gslnretainLRgrav}
g(r) = K \sqrt{rm(r)} 
\end{equation}
This is the geometric mean of the two lengths $r$ and $m(r)$ multiplied by an arbitrary constant $K$. This is the precise form of $g(r)$ which makes the sound speed constant radially, see Eq. \eqref{eq:enth}. If this form of $g(r)$ can be engineered in the lab, the now constant $c_s^2$ factor can be absorbed by a proper coordinate transformation $dt \to dt_{\textsc{ac}} = c_sdt$ and the analogy between the metrics Eqs. \eqref{eq:metrtac} and  \eqref{eq:PGmetric} becomes exact, and, among other things the light ring structure of the gravitational target is retained.

If we want to be less restrictive and are interested only in maintaining intact the precise positions of light rings but not the entire lapse structure, we need to impose the relation \eqref{eq:retainLRgrav} between $g'$ and $m'$ only at the positions $r=r_{\textsc{lr}}$. It is now an algebraic relation between two numbers and not a differential equation to solve, hence the solution \eqref{eq:gslnretainLRgrav} is only one among an infinite set of functions that pass through the right values of $m'$ and $g'$ at $r=r_{\textsc{lr}}$.

For the sake of this analytic analysis, we stick to the more restrictive case of $\sigma = 0$ at all $r$ and report the necessary behavior of $g(r)$ in the two limiting regions $r\to 0$ and $r\to\infty$, where $m(r) \to r^3/2\ell^2$ and $m(r) \to M$ respectively

\begin{align}
\label{eq:glimitbehav}
g(r) \to \frac{K}{\sqrt2\ell}r^{2} \quad &{\rm for} \, \, r\to 0\\
g(r) \to K\sqrt{M}r^{1/2} \quad &{\rm for} \, \, r\to \infty \nonumber
\end{align}

It is worth stressing that it is the full prescription \eqref{eq:gslnretainLRgrav}, and not either of its asymptotic power laws \eqref{eq:glimitbehav}, that cancels $\sigma$ everywhere: imposing the outer form $g\propto r^{1/2}$ globally recovers most of the displacement but not all of it, leaving the two-light-ring window at $0.77\lesssim\ell/M\lesssim0.93$ instead of the gravitational $0.95$.

Engineering $g(r)$ like in Eq. \eqref{eq:gslnretainLRgrav} and \eqref{eq:glimitbehav} can be done via a radially varying magnetic field through the Feshbach resonance connection between an external magnetic field and the interaction between the cold atoms. We provide an analysis in Appendix \ref{sec:Bfield}. Whether this is experimentally viable is another question, to which Sec. \ref{sec:exp} is dedicated. If this engineering is done, then the analogue spacetime reproduces faithfully \emph{all} features -- regular core, trapped region, horizon structure, light rings, surface gravities -- of the gravitational spacetime. If not, then the lapse function is different and, among other consequences, the light ring structure is shifted as in Eq. \eqref{eq:lwindow} which must be kept in mind, but the simulated object is still a valid regular ultracompact object that we can work with and whose instabilities we can study.

\subsubsection{Global realization of RBH/BHM spacetimes}
\label{subsec:global}

We have been concentrating on simulating RBH/BHM spacetimes mostly in the central and the asymptotic region because it is more difficult to conceive of precisely simulating the entire $m(r) = \frac{Mr^3}{r^3+2Ml^2}$, to use the Hayward model as an example. Furthermore, in the two asymptotic regimes the balance in Eq.~\eqref{eq:ubec} is dominated by a single term and the result can be displayed analytically. What is strictly necessary is to realize the central region, because this is where the inner light ring and inner horizon instabilities originate. The asymptotic region, as we have seen, is simple to realize in the form $m(r)\to M$, but any other behavior that has a sufficiently small curvature would not ruin echo and QNM observations at the sufficiently far away experiment boundary $r_{\rm cb} \gtrsim M$. However, the intermediate region between the two, which we have not analyzed separately from the limiting regions, is also necessary, since this is where the trapped region and the outer horizon live, and they play a role in ensuring the correct causal structure, in which the instabilities of interest play out. 

Nevertheless, it is worth stressing that Eq.~\eqref{eq:ubec} is \emph{not} an approximation: given any mass function $m(r)$ and any choice of $g(r)$, it returns in closed form the external potential that realizes that spacetime at \emph{every} radius, the density and velocity profiles following from \eqref{eq: dens} and \eqref{eq:correq}. No matching or interpolation between an inner and an outer solution is required, and the intermediate region is obtained at no extra cost. 

\begin{figure}
\includegraphics[width=15cm]{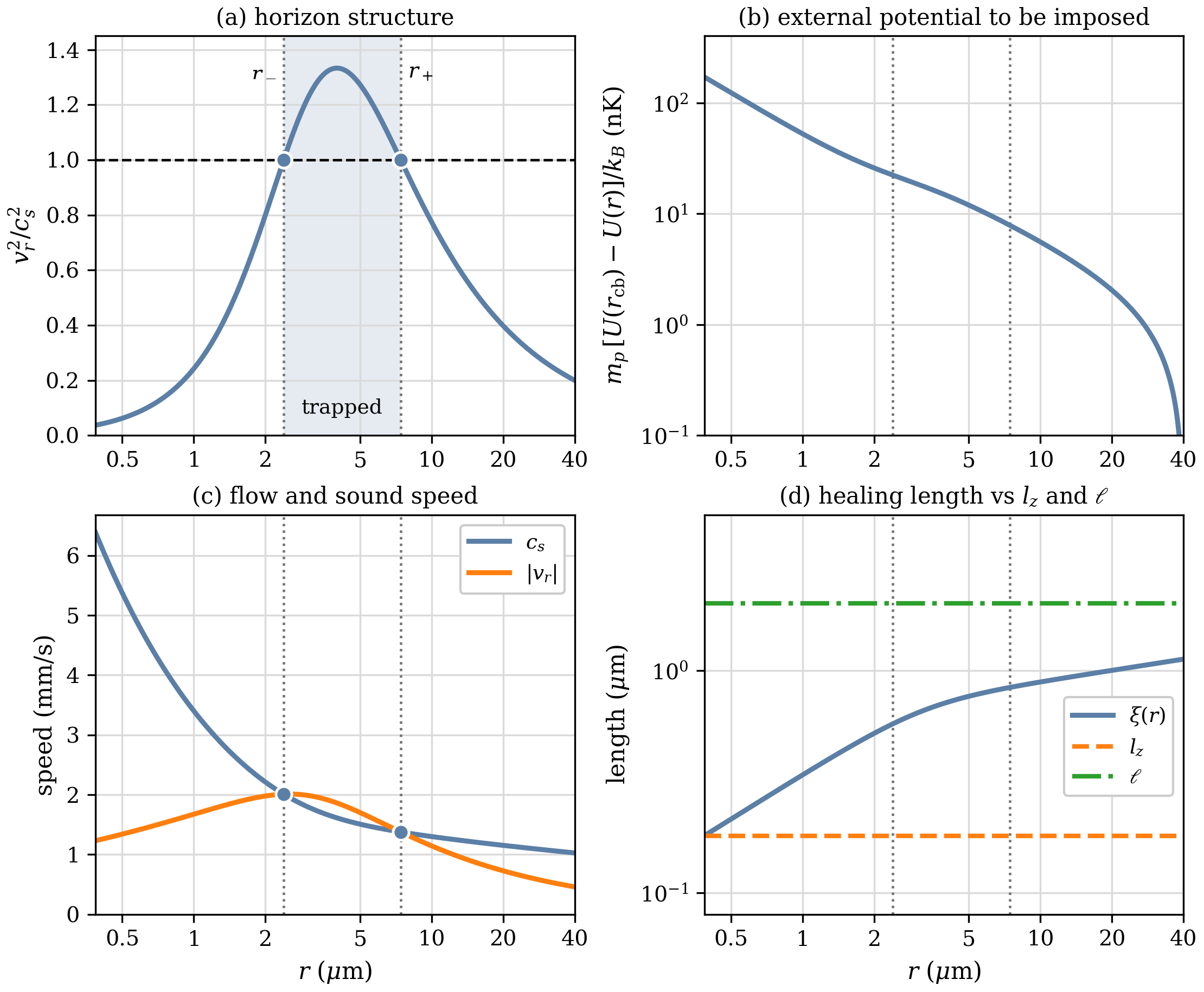}
\caption{The complete solution for the second analog regular black hole of
Table~\ref{tab:designs} (second column): a Hayward spacetime with $M=4\,\mu$m,
$\ell=2\,\mu$m realized at constant $g$ in a container of radius
$r_{\rm cb}=40\,\mu$m, for $r_{\rm min}\le r\le r_{\rm cb}$ with
$r_{\rm min}\simeq0.39\,\mu$m the radius at which $\xi=l_z$ and the condensate
ceases to be quasi-two-dimensional. The profiles follow from Eqs.~\eqref{eq:radvel},
\eqref{eq: dens}, \eqref{eq:enth} and \eqref{eq:ubec}, with the full quantum pressure
of Eq.~\eqref{eq:qp} (for $g'=0$). (a) The flow-to-sound-speed ratio, with the two
horizons and the trapped region between them; (b) the external potential to be
imposed, relative to its value at the wall; (c) sound speed and inflow velocity,
crossing at $r_\pm$; (d) the healing length compared with the transverse
oscillator length $l_z$ and the regularization scale $\ell$. Dotted vertical
lines mark $r_\pm$.}
\label{fig:global}
\end{figure}

Figure~\ref{fig:global} shows this global solution for the analog regular black hole of Sec.~\ref{sec:exp}, a Hayward object with $\ell/M = 0.5$. Panel (a) is the laboratory counterpart of Fig.~\ref{fig:vcssq}: both horizons sit comfortably inside the container, with the trapped supersonic region between them. Panel (b) is the profile that the digital micromirror device must imprint, a smooth and monotonic well of depth a few hundred nK relative to the boundary. Panels (c) and (d) show that the flow is transonic and that the healing length stays between the transverse oscillator length $l_z$ and the regularization scale $\ell$ throughout, which are precisely the conditions collected in Sec.~\ref{subsec:validity}.

On the contrary, if it is not possible to realize precisely one of the exact theoretical models $m(r)$ that exist, the efforts to simulate regular UCOs and observe their instabilities can be dictated my experiments. Provided that the central region is regularized, an inner and outer horizon or inner light ring are present in the analog spacetime, and some equivalents of the regularization parameter and the ADM mass exist, we have at hand a valid regular static, spherically symmetric spacetime. If experimentalists are able to realize any sort of such regular UCO spacetime and observe radiation at infinity, theorists could then extract from this signal the corresponding $m(r)$. This is a more experiment-dictated bottom-up approach, in contrast to the theory-dictated top-down approach in the present work, which was to impose an $m(r)$ model and derive the necessary experimental setup to realize it.

\section{Experimental considerations}
\label{sec:exp}

\subsection{The stationary experimental setup}
\label{subsec:setup}
We propose an analog experimental platform based on a quasi two-dimensional BEC of atoms of potassium-39 of mass $m_p \approx 6.47 \times 10^{-26}$ kg, cooled to nK temperatures via consecutive laser and evaporative cooling. The dimensional reduction occurs thanks to a tight confinement of the atoms in the vertical direction with a harmonic dipole trap, which effectively limits the dynamics to a disc in the horizontal plane. The external potential per mass $U(r)$ is used to radially confine the atoms to a certain confinement boundary radius $r_{\rm cb}$ and to impose a certain velocity and density profile. The potential is controlled via a digital micromirror device (DMD), which allows for considerable freedom in choosing the exact shape $U(r)$.  

Imaging can be performed via standard absorption imaging techniques and $c_s(r)$ will be measured by photon excitation and consecutive wave front tracking. The atomic interaction strength $g(r)$ entering the sound speed can be tuned via the Feshbach resonance. The magnetic field can be realized using two coils, hence the simplest behavior will be the expansion $B(r) \approx B_0 + const  \cdot r^2$, resulting in a tunable in size, but radially constant $g$ if one stops at the first term, or a tunable and radially varying $g(r) = \tilde{g}\left(\frac{\tilde{r}}{r}\right)^a$ with $a=2$ if one takes also the second term. Since the atoms possess a magnetic moment $\mathbf{d}$, a magnetic field gradient would cause an additional force on them which must be compensated for by an extra term in $U(r)$ equal to $-\mathbf{B} \cdot \mathbf{d}$. 

In Appendix \ref{app:varg} we show the calculation steps to estimate $\ell$ and $M$ for such an experimental setup and evaluate the feasibility of the easiest to realize non-constant $g(r) = \tilde{g}\left(\frac{\tilde{r}}{r}\right)^2$, as it is instructive to see why precisely it fails to be viable. It is for this reason that the designs we settle on in Sec.~\ref{subsec:designs} will use a spatially uniform $g$ in the end, which we already mentioned in Appendix~\ref{sec:Bfield}, but the precise experimental reason will now be carefully spelled out in Sec.~\ref{subsec:validity}. The magnetic field is then a constant and the dipole compensation term is absent, which is a considerable simplification of the setup described above and yet another reason to prefer a constant $g$. A residual, weak radial variation remains available as a tuning knob, at the price of reinstating the compensation.

\subsection{Behavior at the center and alternative configurations}

Since we require a stationary inflow of condensate towards the center and the three variables $(\rho(r), v_r(r), U(r))$ formally diverge at $r=0$, the question arises what happens at the center. It is clear that we do not truly want or need divergent quantities, and are interested not in the precise center, but in the core region $r\ll M$. This is the region where we require the behaviors of the four variables derived in the previous section. This leaves the behavior at the precise center $r=0$ free to be determined by what is experimentally most viable and simple to realize. 

One option that is used in analog-gravity platforms with an inflow is to place a sink for the condensate in the center and refill it somewhere else, as is often done in superfluid Helium experiments \cite{Smaniotto_2025, Svancara_2024}, because otherwise stationarity will be lost due to the loss of particles and, as a result, a time dependence of density, velocity and sound speed will be introduced. In BEC experiments, the drainage can be performed but not the refilling\footnote{A BEC is an isolated, vacuum-trapped coherent sample with no inlet to an external reservoir, so new atoms would have to be freshly cooled and phase-matched to the existing flow, unlike the macroscopic plumbing used to recirculate liquid helium in Refs.~\cite{Smaniotto_2025,Svancara_2024}. Removing
atoms at the center, however, is dissipative rather than coherent and is routinely done via RF/Raman outcoupling~\cite{Mewes:1997zz, PhysRevLett.82.3008} or localized engineered loss~\cite{Barontini2013}, either of which can supply the sink required by the stationary inflow of Eq.~\eqref{eq:radvel}.}, so alternatives must also be considered to ensure stationarity.

An alternative option could be an oscillating configuration, where $\rho$ becomes higher and lower periodically and $v_r$ alternates between inflow and outflow in an approximately harmonic (sinusoidal) manner. In the almost-stationary phases of this oscillation, the condensate could be sufficiently similar to the stationary configuration necessary for our simulation. The almost-stationary phases need to be long enough to be able to see the instabilities, see Eq.~(\ref{eq:tauresults}) for the instability timescales. On the other hand, if they are too long, one could start seeing also the black hole laser instability predicted by \cite{Corley_1999,Finazzi:2010nc} for systems with modified dispersion relation of the type
\begin{equation}
\label{eq:moddisprel}
    \omega^2 = c_s^2k^2 + k^4/k_0^2 \, \, ,
\end{equation}
see ref.~\cite{Corley_1999} Eq.~(2.7). This is exactly the dispersion relation of a BEC with healing length $\xi$, if one identifies $k_0^2 = 4/\xi^2$. Indeed, this is one of many reasons to use BECs as an analogue platform, which, due to their dispersive deviations from relativistic wave propagation at high frequencies, allow for testing Lorentz-breaking quantum gravity effects expected in effective field theory frameworks. It would be interesting to analyze the fate of the inner horizon or inner light ring instabilities under these modifications of the dispersion relation and we leave this for future investigations.

While the black hole laser instability could be a nuisance for the present work's objective of observing the inner horizon or inner light ring instability, observing it would be great progress for the analog gravity field and thus merits detailed considerations on its own that we delegate to future work.

In conclusion, while in this paper we have focused on stationary configurations because they allow for simple analytic computations and considerations, experimental efforts may be more fruitful if directed at oscillating configurations. Alternatively, dynamic situations are studied semi-analytically and numerically in the literature and are in no way less interesting, see Appendix~\ref{app:dyn} for an example. Experimental attention should be turned towards them as well, since they are physically more viable (static spacetimes being an astrophysically unrealistic idealization), and even more so if they prove to be easier to realize in the lab than the present stationary configuration.

Other configurations are effective 1+1-dimensional, ring-like stationary setups like in \cite{Garay_2000, Garay_2001} are also currently experimentally viable, provided the recycled modes due to the periodic boundary conditions of the problem are sufficiently damped. These setups deserve a dedicated analysis that we leave to future work. 

\subsection{Validity conditions and the feasible region}
\label{subsec:validity}

The analysis in App. \ref{app:varg} suggests organizing the remaining discussion as follows. Rather than fixing experimental values and reading off $\ell$ and $M$, one should first write down the conditions that any realization must satisfy, and then look for the region of parameter space in which all of them. In the present subsection, we list the conditions in a form in which each is a single inequality between laboratory quantities. In the next subsection, \ref{subsec:designs}, we will propose three fiducial designs that satisfy them, presenting $\ell$ and $M$ for them as calculated with the steps in the Appendix. Finally, in \ref{sec:instab}, we calculate the resulting estimates for the instability timescales.

\paragraph*{(i) The geometry must fit inside the container.} The features to be studied, and enough of the asymptotic region beyond them to let modes propagate out and be detected, must be contained well within $r_{\rm cb}$. For a RBH this means $r_+\simeq 2M\ll r_{\rm cb}$, for a BHM that the outer light ring of Eq.~\eqref{eq:LRanalog}, at $r_\textsc{olr} \simeq2.2$--$2.4M$, must be. Calling $r_{\rm out} \equiv r_+\, \, {\rm or} \, \, r_\textsc{olr}$ for RBH or BHM respectively, in practice we impose $r_{\rm cb}\gtrsim 4$--$5\,r_{\rm out}$.

\paragraph*{(ii) The condensate must remain quasi-two-dimensional.} This requires $m_pc_s^2(r)/2\lesssim\hbar\omega_{\rm z}$ everywhere, in order to avoid excitement above the transverse harmonic ground state. Using $\xi = \hbar/(\sqrt{2 m_p c_s})$ and $l_z = \sqrt{\hbar/m_p\omega_{\rm z}}$ this takes the remarkably simple form
\begin{equation}
\label{eq:quasi2d}
\frac{m_pc_s^2(r)/2}{\hbar\omega_{\rm z}} = \frac{l_z^2}{4\xi^2(r)}\lesssim1
\qquad\Longleftrightarrow\qquad
\xi(r)\gtrsim l_z\,.
\end{equation}
It is worth noting, for the two-dimensional coupling $g=\sqrt{8\pi\hbar^3\omega_{\rm z}/m_p}\,a$, that the ratio is fixed by a single dimensionless combination,
\begin{equation}
\label{eq:xioverlz}
\frac{\xi}{l_z}=\left[2\sqrt{8\pi}\;a\,n_{2D}\,l_z\right]^{-1/2}\,,
\end{equation}
so that a dilute, weakly interacting, tightly confined cloud is what one wants. One also needs $a\ll l_z$ for that coupling to be valid in the first place.

\paragraph*{(iii) The features must be resolved.} The hydrodynamic and eikonal descriptions require the healing length to be small compared with the scale of the structure being probed: $\xi(r)\ll\ell$ and $\xi(r)\ll r_-$ (respectively $\xi\ll (r_{\textsc{olr}}-r_{\textsc{ilr}})$ for the light-ring pair, to resolve the two light ring positions). Together with (ii) this sandwiches the healing length,
\begin{equation}
\label{eq:sandwich}
l_z\;\lesssim\;\xi(r)\;\ll\;\min\left(\ell,\,r\right)\,,
\end{equation}
and, since $c_s^2\propto (rm)^{-1/3}$ grows ($\xi(r)$ reduces) towards the center, the left inequality together with the right one bind in the core at larger radii. The same thing happens as the interaction strength increases. Condition \eqref{eq:sandwich} is the sharpest way to state what an analog UCO needs: it can be met only if $\ell \gg l_z$, i.e.\ if the regularization scale is large compared with the transverse oscillator length, which is why tight vertical confinement is an asset. Note also that \eqref{eq:sandwich} guarantees $h\gg|Q|$ through Eq.~\eqref{eq:hoverQ}, so no fine-tuning of $g(r)$ is needed to keep the quantum pressure subdominant --- it already will be based on these conditions.

\paragraph*{(iv) The sample must live long enough against three-body loss.} With $n_{3D}\simeq n_{2D}/(\sqrt{2\pi}\,l_z)$ and a recombination coefficient $L_3\simeq 3C\hbar a^4/m_p$ with $C=\mathcal{O}(10^2)$ away from Efimov features \cite{Soding_1999, Etrych_2023}, the requirement is $\left(L_3n_{3D}^2\right)^{-1}\gg\tau$, with $\tau$ the instability or echo timescale of interest. Because of the $a^4$ scaling this is the condition that most sharply excludes near-resonant interaction strengths.

\paragraph*{(v) The stationary inflow must last long enough.} The configuration we initially assume drains condensate through the center at a rate $F = 2\pi n_{2D}(r)|v_r(r)|r$, constant by continuity, and the whole disc contains only $N\simeq \pi r_{\rm cb}^2 n_{2D}$ atoms. The condensate therefore survives for $t_{\rm drain}=N/F$, and one needs $t_{\rm drain}\gg\tau$\footnote{A very large difference between those two time scales is required because much before $t_{\rm drain}$ the loss of atoms will induce a time dependence on the fluid quantities, breaking stationarity and changing the background geometry.}. Using $|v_r(r_{\rm cb})|=\sqrt{2M/r_{\rm cb}}\,c_s$ and $\tau_{\textsc{ilr}}\sim4M/c_s$ one finds the useful scaling
\begin{equation}
\label{eq:drain}
\frac{t_{\rm drain}}{\tau_{\textsc{ilr}}}\;\sim\;\frac{1}{8\sqrt2}\left(\frac{r_{\rm cb}}{M}\right)^{3/2}\,,
\end{equation}
so that a large container helps twice over, through (i) and through (v). This condition is a genuine limitation of the stationary setup and is one more argument in favor of the oscillatory or ring setups discussed in Sec.~\ref{subsec:setup} or the step-wise stationary configurations illustrated in Appendix~\ref{app:dyn}. These must fulfill an analogous experimental condition that the time the system remains in the almost-stationary phase must also be sufficiently larger than the instability timescale. In the oscillatory setups, this is the time the setups remains in the almost-stationary part of the oscillation (close to an oscillation extremum); in ring setups, it is the time before recycled modes in the ring have caused an instability; in step-wise stationary configurations, it is the duration of the step-wise stationary phase.

\paragraph*{(vi) The potential must be realizable.} The depth of $U(r)$ over the simulated region, in temperature units $m_p|U|/k_B$, must be within the reach of a DMD-shaped in-plane potential, i.e.\ at most a few hundred nK for a cloud at these densities.

Conditions (ii), (iv) and (vi) all push in the same direction --- towards smaller $a$ and smaller $n_{2D}$ --- while (iii) pushes the other way, and (i) and (v) are satisfied by making $r_{\rm cb}/M$ large. The resulting window is narrow but, as we now show, not empty.

\subsection{Three fiducial designs}
\label{subsec:designs}

Table~\ref{tab:designs} collects one parameter set for two analog RBH and one analog BHM which satisfy all of (i)--(vi) simultaneously, with quasi-2D potassium-39 parameters of the kind already demonstrated in \cite{Viermann_2022, Tolosa_Sime_n_2022, duchene2026maxwellfisheyelensboseeinstein}. All use a spatially uniform interaction strength. The complete profiles for the second RBH entry are the ones plotted in Fig.~\ref{fig:global}.

Several features of the table deserve comment. The interaction strength is small, $a=30\,a_0$, and the areal densities a few times $10^{12}\,{\rm m^{-2}}$, an order of magnitude below the values in Table~\ref{tab:expval}; this is what buys a healing length comfortably above $l_z$ while keeping the three-body lifetime in the hundreds of seconds. Tight vertical confinement, $\omega_{\rm z}=3\times10^4$--$10^5\,{\rm rad/s}$, is what makes the sandwich \eqref{eq:sandwich} wide enough to be usable. The required potentials are of order $10^2$ nK deep, well within reach.

The first RBH entry is comfortable: every validity condition is met with an order of magnitude to spare. The trapped region sits well inside the container, $\tau_{\textsc{IH}}\approx6$ ms against a drain time of $\approx160$ ms. This is ample --- not the lifetime of the sample, but the necessity for taking into account modified dispersivity and full nonlinear backreaction is what will limit immediate applicability. For the BHM entry, there are two numbers within the table that show that it is sitting very tightly within the realizable regime with virtually no margin. The first is $(r_{\rm out}-r_{\rm in})/\xi(r_{\rm in}) = 1.1$, meaning that the two light rings are barely resolved. The second number is $t_{\rm drain}/\tau_{\rm echo}\approx3$. Indeed, the echo time delay between the two light rings, used as a geometric estimate\footnote{To be precise, the echo time delay is not an exact quantification of the nonlinear instability timescale. However, the latter is highly model-dependent and there is no analytic formula available in the literature. We have checked that the echo delay yields results not far off from literature estimates. For example, a constraint derived by \cite{Cunha_2023} is $\tau_{\rm ILR} \lesssim 10^3  \, \frac{4M}{c_s}$.} of the time frame within which a linear instability could start manifesting if a continuous influx of perturbations is provided, is estimated by\footnote{This is the dominant contribution in the echo time delay for ultracompact stars of a given compactness, see \cite{Arrechea_2024}. It is the null-ray bounce time between the regular center, acting as a centrifugal reflective barrier, and the outer light ring, a leaky wall through which the signal reaches an outside observer. The lower limit $r\to0$ is again not to be taken literally, but in any case, the
integral is dominated by the region around $r_{\rm in}$, where $f(r)$ approaches its minimum, which in the BHM case is non-zero.}
\begin{equation}
\label{eq:echodelay}
\tau_{\rm echo} = 2\int_{0}^{r_{\textsc{olr}}}\frac{dr}{c_s(r)f(r)}\,.
\end{equation}
\val{}It grows without bound as $\ell\to\ell_*$ from above, because the object then approaches the threshold of horizon formation and $f$ approaches zero; at $\ell/M=0.79$, for example, one finds $\tau_{\rm echo}\approx130$ ms, longer than the drain time, and no echo would be seen. Increasing $\ell/M$ shortens the delay, but Eq.~\eqref{eq:lwindow} forbids going beyond $\ell/M\simeq0.895$, where the analog light rings merge. The BHM operating window of this platform is therefore squeezed between these two requirements,
\begin{equation}
\label{eq:bhmwindow}
0.82\;\lesssim\;\ell/M\;\lesssim\;0.895\,,
\end{equation}
and the entry in Table~\ref{tab:designs}, at $\ell/M=0.88$, sits inside it and yields around three echoes within the drain time. That is enough to identify an echo train and to measure its delay, and thus to test the compactness--delay relation; it is \emph{not} enough to follow the non-linear development of the light-ring instability, which would require the oscillatory configurations or ring setups of Sec.~\ref{subsec:setup} or the dynamical setups of Appendix~\ref{app:dyn}, or a substantially larger disc but also considerations on backreaction.

We explore a second RBH design (the second column of Table \ref{tab:designs} to maintain the experiment size $r_{\rm cb}=40 \,\mu$m, since this is a parameter that is difficult to modify in an already-existing experiment. The two regular-black-hole designs are related by a rescaling of the target
geometry by $1/2$, which keeps $r_{\rm cb}/M=10$ and $t_{\rm drain}/\tau_{\rm IH}\simeq25$ fixed. Keeping $r_{\rm cb}=40\,\mu$m instead costs margin in the sandwich Eq. \eqref{eq:sandwich}: to maintain $\xi\ll\ell$ at a smaller $\ell$ we raise the density and lower
$\omega_z$, which pushes $r_{\rm min}$ out to $\simeq0.19\,\ell$ (instead of
$0.065\,\ell$). The inner horizon at $r_-\simeq1.2\,\ell$ and the outer part of the de Sitter core are resolved, but the deep core is not. Furthermore, $\xi/l_z=3.2$ and $\xi/\ell=0.29$ at
$r_-$ against $6.5$ and $0.21$ before, and the trapped region now spanning only
$8.8$ healing lengths rather than $12.0$. Conditions (ii)--(iii) are thus tightened, but every condition is still satisfied with a margin of around an order of magnitude. The smaller design is thus a viable substitute maintaining the size of an already existing experiment, hence why it was chosen as a prototype to illustrate in Fig. \ref{fig:global}.

\begin{table*}[t]
\centering
\begin{tabular}{l c c c}
\hline\hline
 & analog RBH & analog RBH & analog BHM \\
 & ($r_{\rm cb}=80\,\mu$m) & ($r_{\rm cb}=40\,\mu$m) & ($r_{\rm cb}=40\,\mu$m) \\
\hline
\multicolumn{4}{l}{\emph{target geometry (Hayward)}} \\
ADM mass $M$                        & $8\,\mu$m  & $4\,\mu$m   & $2\,\mu$m \\
regularization scale $\ell$         & $4\,\mu$m  & $2\,\mu$m   & $1.8\,\mu$m \\
$\ell/M$                            & $0.50$     & $0.50$      & $0.88$ \\
features ($\mu$m)                   & $r_\pm=4.8,\,14.8$ & $r_\pm=2.4,\,7.4$ & $r_{\textsc{lr}}=3.5,\,4.3$ \\
\hline
\multicolumn{4}{l}{\emph{condensate}} \\
container radius $r_{\rm cb}$       & $80\,\mu$m & $40\,\mu$m  & $40\,\mu$m \\
vertical trapping $\omega_z$        & $10^{5}\,$rad/s & $5\times10^{4}\,$rad/s & $6\times10^{4}\,$rad/s \\
oscillator length $l_z$             & $0.13\,\mu$m & $0.18\,\mu$m & $0.17\,\mu$m \\
scattering length $a$ (uniform)     & $30\,a_0$  & $30\,a_0$   & $30\,a_0$ \\
areal density at $r_{\rm cb}$       & $3\times10^{12}\,{\rm m}^{-2}$ & $9\times10^{12}\,{\rm m}^{-2}$ & $8\times10^{12}\,{\rm m}^{-2}$ \\
atom number $N$                     & $7\times10^{4}$ & $5.5\times10^{4}$ & $4.9\times10^{4}$ \\
$r_{\rm min}$ (where $\xi=l_z$)     & $0.26\,\mu$m & $0.39\,\mu$m & $0.26\,\mu$m \\
\hline
\multicolumn{4}{l}{\emph{validity checks}} \\
(i) $r_{\rm cb}/r_{\rm out}$, $r_{\rm out} = r_+$ or $r_{\textsc{olr}}$                         & $5.4$   & $5.4$   & $9.3$ \\
(ii) $\xi/l_z$ at the feature (at $r_{\rm in}$)                        & $6.5$   & $3.2$   & $4.4$ \\
(ii) $a/l_z$                                          & $0.012$ & $0.009$ & $0.010$ \\
(iii) $\xi/\ell$ at the feature                       & $0.21$  & $0.29$  & $0.41$ \\
(iii) $(r_{\rm out}-r_{\rm in})/\xi(r_{\rm in})$         & $12.0$  & $8.8$   & $1.1$ \\
(iii) $r_{\rm in}/\xi(r_{\rm in})$                        & $5.7$   & $4.2$   & $4.8$ \\
(iv) three-body lifetime at the feature
                                                      & $250\,$s & $56\,$s & $143\,$s \\
(v) drain time $t_{\rm drain}$                       & $156\,$ms & $53\,$ms & $76\,$ms \\
(vi) $m_p{\rm max}(|U|(r))/k_B$ over $r_{\rm min}\le r\le r_{\rm cb}$ & $340\,$nK & $170\,$nK & $186\,$nK \\
\hline
\multicolumn{4}{l}{\emph{observables}} \\
$c_s$ at the feature                & $1.4\,$mm/s & $2.0\,$mm/s & $1.6\,$mm/s \\
$\kappa_-$                          & $0.12\,\mu{\rm m}^{-1}$ & $0.23\,\mu{\rm m}^{-1}$ & --- \\
timescale $\tau$                    & $\tau_{\rm IH}=6.3\,$ms & $\tau_{\rm IH}=2.2\,$ms & $\tau_{\rm echo}=26.2\,$ms \\
$t_{\rm drain}/\tau$                & $25$ & $25$ & $2.9$ \\
\hline\hline
\end{tabular}
\caption{Three parameter sets satisfying all the conditions (i)--(vi) of
Sec.~\ref{subsec:validity}: two realizing a regular black hole with two horizons
and one a horizonless mimicker with two light rings. All designs use
a spatially uniform interaction strength. ``Feature'' denotes the inner horizon
$r_-$ for the RBHs and the inner light ring $r_{\textsc{ilr}}$, for the BHM, both collectively denoted by $r_{\rm in}$, while the outer horizon
$r_+$ for the RBHs and the outer light ring $r_{\textsc{olr}}$, for the BHM are collectively written as $r_{\rm out}$. Light-ring positions are
those of the analog geometry, Eq.~\eqref{eq:LRanalog}; the gravitational
positions for the same $\ell/M$ would be $6.5$ and $9.9\,\mu$m. The potential
depth (vi) is evaluated over the region $r\ge r_{\rm min}$ in which the cloud is
quasi-two-dimensional, $\xi\ge l_z$, and the three-body lifetime (iv) uses
$L_3=3C\hbar a^4/m_p$ with $C=10^2$ at the density of the condensate at the
feature. The complete radial profiles of the second column are shown in
Fig.~\ref{fig:global}.}
\label{tab:designs}
\end{table*}

\subsection{Numerical estimate of the instability timescales}
\label{sec:instab}

With the results for $\ell$ and $M$, we can estimate the inner light ring and the inner horizon instability timescales. As said, as a kinematic estimate for the former we use the null-ray bounce time between the effective potential walls \eqref{eq:echodelay}.

For the latter, the timescale is~\cite{Carballo-Rubio:2018pmi, Carballo-Rubio:2021bpr, DiFilippo:2022qkl}
\begin{equation}
\label{eq:tauih}
\tau_{\rm IH} \sim \frac{1}{c_s(r_-)\,|\kappa_-|} \quad {\rm with} \quad \kappa_- = \frac{1}{2}\frac{d}{dr}\left.\left(1-\frac{2m(r)}{r}\right)\right|_{r_-}
\end{equation}
where the factor of $c_s(r_-)$ is exactly the conversion between the analog surface gravity and an inverse laboratory time identified in Sec.~\ref{sec:residual}. The inner horizon position is the smallest positive root of $r^3-2Mr^2+2M\ell^2=0$, i.e. \cite{Perez_Roman_2018}
\begin{equation}
\label{eq:rminus}
    r_- = \frac{2M}{3} + \frac{4M}{3}{\rm cos}\left[\frac{\pi}{3} + \frac{1}{3}{\rm arccos}\left(\frac{27\ell^2}{8M^2}-1 \right)\right] \,.
\end{equation}
Thus, for the three designs of Table~\ref{tab:designs} we obtain
\begin{align}
\label{eq:tauresults}
\tau_{\textsc{ih}} &\sim 6\times10^{-3} \, {\rm s} \qquad\text{(1. analog RBH)},\\
\tau_{\textsc{ih}} &\sim 2\times10^{-3} \, {\rm s} \qquad\text{(2. analog RBH)},\\
\tau_{\rm echo} &\sim 26\times10^{-3} \, {\rm s} \qquad\text{(analog BHM)},
\end{align}
These fall within a well measurable timeframe, given that as a reasonable unit of time in the experiment we can take $(\omega_{\rm z}/2\pi)^{-1} = \mathcal{O}({\rm ms})$, and, by construction, within the drain time \eqref{eq:drain}. Since under slight changes of experimental values, these instability time scales can also change significantly, it is possible that only one of these instabilities' timescales could be made to fall within a reasonably measurable range. This would mean that only a RBH or only a BHM would be realistic to emulate in this analog-gravity platform, and this would still be good progress for the field.

\section{Discussion and conclusions}
\label{sec:concl}
We have shown how to emulate a static, spherically symmetric regular ultracompact spacetime in a quasi-two-dimensional Bose--Einstein condensate --- not merely a horizon or an ergoregion, but the geometry in its entirety. It is done by establishing a mapping between the acoustic and gravitational line elements on an equatorial disc, proposed in \cite{Visser_2005} for the Kerr slice and to our knowledge never revisited since. The mapping is exact as far as the causal structure is concerned: horizons, trapped region and surface gravities are reproduced without approximation. A radially varying sound speed does leave a residual factor $c_s^2(r)$ in the analog $g_{tt}$, which no coordinate or conformal transformation can remove and which displaces the null circular orbits. The resulting two-light-ring window, Eq.~\eqref{eq:lwindow}, is narrower or wider than the gravitational one depending on the interaction strength $g(r)$ and must be imposed on the analog geometry when designing an experiment --- unless $g(r)$ is engineered as in Eq.~\eqref{eq:gslnretainLRgrav}, in which case the displacement cancels identically. Everything else follows from Eq.~\eqref{eq:ubec}, which fixes the external potential at every radius for any $m(r)$, so that the whole spacetime, and not merely its core and its asymptotics, is realized at once (Fig.~\ref{fig:global}).

The main message of this work is that a condensate is a promising platform for simulating black holes possessing \emph{both} an outer \emph{and} an inner horizon. This could bring within laboratory reach the instabilities that afflict Cauchy horizons --- the classical mass-inflation growth of the internal mass parameter, and the semiclassical divergence of the renormalized stress tensor in the Unruh vacuum, which requires precisely $|\kappa_-|\neq|\kappa_+|$. The two RBH designs of Table~\ref{tab:designs} place the entire trapped region well inside the container, spanning some ten healing lengths, with $|\kappa_-|/|\kappa_+|\approx4$ and a draining time that surpasses the inner-horizon instability timescale by a factor $\approx25$. This could be a viable way of ensuring stationarity of the analog geometry for long enough to observe the instabilities before the inflow is exhausted, \emph{if} a proper refilling mechanism is successfully implemented --- if that proves difficult for atoms that must be kept ultracold, alternatives like an oscillating or a 1+1-dimensional ring configuration must be explored. A condensate moreover offers the opportunity to test modified-gravity phenomenological effects: its superluminal dispersion allows the blueshifted modes to escape the trapped region, so that a signature of the inner-horizon instability could become visible at infinity as an ultraviolet excess over the Hawking spectrum, in frequency bands set by $c_s|\kappa_-|$. Whether the instability survives the $k^4$ term --- we expect it tamed to a power law rather than erased \cite{Coutant:2011fz} --- is itself a question this platform is well placed to answer.

Horizonless mimickers could also be realized, albeit with more difficulty. Their operating window is squeezed from below by the echo delay, which diverges as $\ell\to\ell_*$, and from above by the merging of the analog light rings, Eq.~\eqref{eq:bhmwindow}. More severely, the two light rings of our best design sit at $1.1$ healing lengths of each other, at the very edge of what the hydrodynamic description resolves, and the drain time affords only about three echoes. The prize, should these margins be improved or experimentally proven to be sufficient, is access to much of the phenomenology that the gravitational-wave community has developed for horizonless ultracompact objects: echo trains and their compactness-dependent delay, long-lived quasinormal modes, and the conjectured non-linear instability of the inner light ring, whose analog onset one would expect to be signalled by a growing echo amplitude.

Both messages come with the same caveat. The margins are not generous and the estimates depend sensitively on the experimental parameters, so the endeavour to simulate these spacetimes will in practice be dictated as much by what is realistic in the laboratory as by the theoretical construction presented here. The experimental realization, the dynamical and step-wise stationary configurations of Appendix~\ref{app:dyn}, along with oscillating or 1+1-dimensional ring configurations which may well prove easier to realize than the stationary inflow, and the question of backreaction onto the analog geometry, even though it is not governed by the same dynamical equations as general relativity, are all left to future work.

\begin{acknowledgments}
We wish to acknowledge Markus Oberthaler, Francesco Gozzini, David Maibach, Elinor Kath, Jelte Duchêne, Floriane Arrouas, Iacopo Carusotto and Helmut Strobel for insightful discussions. VP acknowledges Markus Oberthaler and the Kirchoff-Institute for Physics of Heidelberg University for the hospitality and the opportunity to visit the ultracold-atoms labs in the final stages of the present work. The authors used ChatGPT (OpenAI) and Claude (Anthropic) as assistive tools for discussing aspects of the physical results, generating the parameter sets in Table \ref{tab:designs} sitting within the feasibility margins of Sec. \ref{subsec:validity}, proof checking and for polishing the manuscript and improving some of the figures. All AI-assisted suggestions were critically assessed, independently checked and approved by the authors, who take full responsibility for the final content.
\end{acknowledgments}

\appendix

\section{A spatially varying $g(r)$ via the Feshbach resonance}
\label{sec:Bfield}

Given this last discussion, it is worth dwelling on how the interaction strength $g(r)$ can be controlled experimentally as a function of radius. The basic machanism behind this is the so called Feshbach resonance, which allows to control $g(r)$ by means of a radially-varying magnetic field as follows \cite{LivingReviews, bagnato2015boseeinsteincondensationyears, Sierant_2017}
\begin{equation}
    g(B(r))=\tilde{g}\left(1-\frac{\Delta}{B(r)-B_0}\right)
\end{equation}
where the constant $\Delta$ is called the resonance width and $B_0$ is the resonance position. Here $\tilde{g}$ is the interaction strength value when the magnetic field is infinitely far away from the resonance position or the resonance width tends to zero. Inverting this relation to see how one needs to engineer the magnetic field to obtain a desired $g(r)$ profile, one obtains
\begin{equation}
    B(r) = B_0 + \frac{\Delta}{1-{g(r)}/{\tilde{g}}}
    \label{eq:reqB}
\end{equation}

We already mentioned that there are two reasons to require a spatially varying interaction strength, which we summarize here:
\begin{enumerate}
    \item Undo the shift in light ring structure caused by the presence of the extra factor of $c_s^2(r)$ in the time-time acoustic metric component, see Eq. \eqref{eq:acdiag}. In this case, $g(r)$ needs to be engineered as in Eq. \eqref{eq:gslnretainLRgrav}, which, in the two limiting regions, becomes Eq. \eqref{eq:glimitbehav}.
    \item To ensure the enthalpy \eqref{eq:enth} is clearly dominant over quantum pressure \eqref{eq:qp} even at arbitrarily small radii $r\to0$ (which, as we mentioned, is not strictly speaking necessary). In this case, one has to engineer $g(r)$ in a way that $h(r)$ rises towards the center more steeply than $|Q(r)|$, i.e.~more steeply than $r^{-2}$. For this, one needs at least $g\propto r^{-1}$, albeit the easiest behaviour to realize experimentally would be $r^{-2}$.
\end{enumerate}

In the first case, the required magnetic field, inserting Eq. \eqref{eq:gslnretainLRgrav}, becomes
\begin{equation}
\label{Bwithgundoshift}
    B(r) = B_0 + \frac{\Delta}{1-\frac{K}{\tilde{g}}\sqrt{rm(r)}}  \, .
\end{equation}
We now clearly see that, since $m(r)$ has dimensions of length, the integration constant $K$ has dimensions of "$g$ over length", hence let us define a length scale $\tilde{R}$ such that $K \equiv \tilde{g}/\tilde{R}$, allowing us to write the following formulas in a more concise way. Using this and inserting Eq.~\eqref{eq:glimitbehav}, we obtain the leading and next-to-leading contributions of the necessary $B(r)$ in the two limiting regimes as
\begin{align}
\label{eq:Blimitbehavundo}
    B(r) \approx (B_0 + \Delta) +\frac{\Delta}{\sqrt{2}}\frac{r^2}{\tilde{R}\ell} 
     \quad &{\rm for} \, \, r\to 0\\
    B(r) = B_0 - \frac{\Delta \tilde{R}}{\sqrt{Mr}}
    \quad &{\rm for} \, \, r\to \infty \nonumber
\end{align}

In the second case, to additionally suppress quantum pressure with respect to enthalpy, one would have to realize $g(r) = \tilde{g}\left({\tilde{r}}/{r}\right)^a$ with $a\geq1$ and $(\tilde{g}\tilde{r}^a)$ practically constant for experimental purposes. The required magnetic field, Eq.~\eqref{eq:reqB}, then becomes
\begin{equation}
\label{eq:Bwithgquantumsuppress}
    B(r) = B_0 + \frac{\Delta}{1-\left({\tilde{r}}/{r}\right)^a}
\end{equation}
which in the two limiting regions takes the form
\begin{align}
\label{eq:Blimitbehavsuppress}
    B(r) \approx B_0-\Delta\left(\frac{r}{\tilde{r}}\right)^a 
     \quad &{\rm for} \, \, r\to 0\\
    B(r) \approx (B_0 + \Delta) + \Delta\left(\frac{\tilde{r}}{r}\right)^a 
    \quad &{\rm for} \, \, r\to \infty \nonumber
\end{align}

We can now understand why $a=2$ was said to be the most experimentally realistic scenario. To realize it, we see that one needs, at the boundary, an almost constant magnetic field (with a small tail falling off like $r^{-2}$), whereas towards the center, it is an inverse parabola. This is a simple series expansion of a trapping magnetic field around a trap minimum $r=0$ and is standard procedure in magnetic trapping of cold atoms \cite{ketterle1999makingprobingunderstandingboseeinstein, gov1999magnetictrappingneutralparticles}. Alternatively, one could attempt to obtain the desired $g(r)$ also with optical resonant spatial modulation of the interaction length.

Let us make two important remarks in view of these results. In both cases Eqs. \eqref{eq:Blimitbehavundo} and \eqref{eq:Blimitbehavsuppress}, the required field diverges where $g(r) = \tilde{g}$, i.e., at $\sqrt{rm(r)}=\tilde{R}$ in the first case, and at $r=\tilde r$ in the second. The expressions above are thus usable only for $r\ll\tilde r, \tilde{R}$ so that the reference radii must be chosen well outside the region to be simulated --- $\tilde r, \tilde{R} \gg r_{\rm cb}$ --- and not at the container wall. 

Furthermore, note how the magnetic field necessary for the two separate purposes is actually, in a certain sense, contrasting: the non-constant contribution has opposite signs in the two cases (not to mention that there is a difference in power in at least one of the regions, depending on the choice of $a$).

We use these remarks to establish a preference a weakly varying or outright constant $g(r)$, which also facilitates the experimentalists greatly. In fact, to conclude, let us restate that, if the experiment remains at scales sufficiently bigger than the healing length by engineering the constant factor in the external potential \eqref{eq:potasymp} and \eqref{eq:potcoregeneral} accordingly, and the shift \eqref{eq:lwindow} in light ring structure is considered, the experimental complications of spatially varying $g(r)$ can be avoided. 

In addition to these simple analytic considerations, we will find much more far-reaching and substantial limitations to the non-constant $g(r)$ approach in Sec.~\ref{sec:exp} focussing on experimental considerations.

\section{Estimate of $\ell$ and $M$ with a radially varying $g(r)$ and feasibility}
\label{app:varg}
In this Appendix, we show the calculation machinery used to make an estimate of the expected values of $\ell$ and $M$ in a realistic BEC analogue, using a general spatially varying $g(r)$. Specifying to a specific profile we then show that the result is outside of the regime of experimental realizability. This motivates the proposal of three fiducial designs in Sec. \ref{subsec:designs} with $g(r) = const$, for which we apply the machinery from this Appendix to obtain $\ell$ and $M$ and also estimate the resulting instability timescales.

\subsection{Experimental values and calculation steps}

We roughly follow \cite{duchene2026maxwellfisheyelensboseeinstein, Viermann_2022, Tolosa_Sime_n_2022, Etrych_2023}'s experimental values, summarized in Table \ref{tab:expval}. Some of the quantities in the table require explanation. For an experiment of size $r_{\rm cb} = 40 \,\mu\mathrm{m}$, we estimate the core region to be at around $r_{\rm cr} = 1 \,\mu\mathrm{m}$, and will check later for consistency whether $r_{\rm cr} \ll M$ as in eq. (\ref{eq:centrdef}). Although in eq. (\ref{eq:potcoreprop}) we found that for our purposes we need an entirely different radial behavior of the potential than the harmonic $r^2$-shape, experimental values are available for the constant factor $\frac{1}{2}\omega_{\rm r}^2$ in front of this shape. This is why we use $\frac{1}{2}\omega_{\rm r}^2 r_{\rm cb}^2$ as an estimate of the numerical value of the potential in the external region regardless of what the actual shape and the dimensions of the constant factor will be, keeping in mind that here we are only interested in rough estimates as a proof of concept.
\begin{table}[h]
\centering
\renewcommand{\arraystretch}{1.4}
\begin{tabular}{lll}
\hline\hline
\textbf{Quantity} & \textbf{Symbol} & \textbf{Experimental value} \\
\hline
Areal density & $n_{2D}$ & $2\times10^{13} \, \mathrm{m}^{-2}$ \quad \\
Chemical potential per mass & $\mu/m_p$ & $3\times10^{-6} \, \mathrm{J/kg}$ \quad \\
Harmonic potential frequency in radial direction& $\omega_{\rm r}$ & $10^2\ \mathrm{rad/s}$ \\
Trapping frequency in vertical direction & $\omega_{\rm z}$ & $10^4\ \mathrm{rad/s}$ \\
System size & $r_{\rm cb}$ & $40\ \mu\mathrm{m}$ \\
Typical radius in core & $r_{\rm cr}$ & $1\ \mu\mathrm{m}$\\
Typical potential in external region & $\frac{1}{2}\omega_{\rm r}^2 r_{\rm cb}^2$ & $8\times10^{-6}$$\ \mathrm{J/kg}$ \\
Typical interaction length in ext. region& $\tilde{a}$ & $200 \, a_0$\\
\hline\hline
\end{tabular}
\caption{Realistic experimental parameters for a quasi-2D BEC of potassium-39 \cite{duchene2026maxwellfisheyelensboseeinstein, Viermann_2022, Tolosa_Sime_n_2022, Etrych_2023}.}
\label{tab:expval}
\end{table}

To estimate $\ell$ from eq. (\ref{eq:potcoregeneral}), we need to estimate the potential and the interaction strength in the inner region. The relation between $g(r)$ and the interaction length $a(r)$ is $g(r) = \sqrt{\frac{8\pi \hbar^3 \omega_{\rm z}}{m_p}}a(r)$ in the case where $g(r)$ is a 2D interaction strength \cite{Tolosa_Sime_n_2022}\footnote{In 3D the relation for the interaction strength is $g_{\rm 3D} = 4\pi\hbar^2a/m_p$. However, when the atoms are tightly confined in the $z$-direction by a strong vertical trapping potential $\omega_zz^2/2$, the interaction constant of the resulting quasi-2D BEC is, indeed, $g(r) = g_{\rm 3D}(r)/\sqrt{2 \pi}l_z$, where $l_z = \sqrt{\hbar/m_p \omega_z}$ is the transverse oscillator length set by the trap.}. This relation propagates to the constants $\tilde{g}$ and $\tilde{a}$. 

To estimate the potential in the inner region from the value $U_{\rm cb} = \frac{1}{2}\omega_{\rm r}^2 r_{\rm cb}^2 - \frac{\mu}{m_p}$ in the outer, we consider the following. We have found that in the asymptotic region, the quantity $U_{\rm cb}$ goes like $r^{-5/3}$. Around the central region, the potential (minus the chemical potential per mass, which however in this case is negligible) increases with decreasing radius like $r^{-8/3}$. Instead of looking for some smooth juncture between these two regimes or computing way more terms to use the full potential profile in Eq. \eqref{eq:ubec}, as an estimate we simply glue them together at the intermediate value $r_{\rm avg} =(r_{\rm cr} + r_{\rm cb})/2$ to obtain the potential at the typical inner radius as $U_{\rm cr} = U_{\rm cb} \left( \frac{r_{\rm avg}}{r_{\rm cr} } \right)^{8/3} \left( \frac{r_{\rm cb}}{r_{\rm avg} } \right)^{5/3}$.

The last step missing to estimate $\ell$ is to elucidate on the physical meaning of the integration constant $B$. For this, we integrate the continuity equation (\ref{eq:cont}) now without skipping steps:

\begin{equation}
    v_r(r)\rho(r)r = \frac{F}{2\pi} = const\, .
\end{equation}
$F$ stands for the constant flux of fluid crossing circles of radius $r$. $F/2\pi$ is then this flux per radian. We estimate $F$ by assuming, for simplicity, that the mass density at $r =r_{\rm cr}$ is $\rho(r_{\rm cr}) \sim n_{2D}m_p$ and the radial velocity is of the order of magnitude of the sound speed $v_r(r_{\rm cr})\sim c_s(r_{\rm cr})$:
\begin{equation}
\label{eq:flux}
    F = 2 \pi n_{2D}m_pc_s(r_{\rm cr})r_{\rm cr}
\end{equation}
Equation (\ref{eq: dens}) follows if we simply identify $B = \frac{1}{2} \left( \frac{Fm_p}{\pi} \right)^{2/3}$. Inserting this into Eq.~(\ref{eq:potcoregeneral}) and setting $r=r_{\rm cr}$, we obtain for $\ell$ using the fiducal values from Table \ref{tab:expval}

\begin{equation}
    \ell \sim \frac{(-U_{\rm cr})^{3/2}m_p^{3}r_{\rm cr}^2}{2^{1/2}B^{3/2}g(r_{\rm cr})} \approx 10 \, \mu\mathrm{m} \,.
\end{equation}
Similarly, to estimate $M$, we turn to eq. (\ref{eq:potasymp}) and set $r=r_{\rm cb}$ with the result

\begin{equation}
    M \sim \frac{B^3g^2(r_{\rm cb})}{(-U_{\rm cb})^3 m_p^6 \,r_{\rm cb}} \approx 40 \, \mu\mathrm{m}
\end{equation}

Note that these expressions are valid for a general $g(r)$ profile. The final numerical values for $\ell$ and $M$, however, result from inserting the shape $g(r) = \tilde{g}\left(\frac{\tilde{r}}{r}\right)^a$ with $a=2$ here as the easiest to implement experimentally. As reference radius $\tilde{r}$ we choose $r_{\rm cb}$, meaning that the interaction strength will be increasing towards the core $r_{\rm cr}$, whence $g(r_{\rm cr}) = \tilde{g} \left( \frac{r_{\rm cb}}{r_{\rm cr}} \right)^{2}$ and $g(r_{\rm cb}) = \tilde{g}$.

\subsection{Viability of the result}

These estimates already deliver the proof of concept we were after: with current experimental values, the analogues of the gravitational parameters come out at reasonable orders of magnitude, comparable to the size of the apparatus. They also show that both $\ell$ and $M$ depend quite sensitively (we see powers up to 4) on the experimental parameters. This can both be a blessing and a curse. On the one hand, there is a large range of technically implementable $\ell$ and $M$ values, achievable even through small variations of the experimental parameters. At the same time, the values that are viable for our purposes must roughly fall within the range $\{r_{\rm cr}, r_{\rm cb}\}$, which is not a particularly large margin and thus easy to fall outside of.

{Taken at face value, in fact, this particular parameter point does \emph{not} define a viable experiment, and each failure points at a condition that we impose systematically in Sec. \ref{subsec:validity}. First, $\ell/M\approx0.25$ places the object on the regular-black-hole branch, so that the inner-light-ring estimate below does not apply to it; and the corresponding horizons, the roots of $r^3-2Mr^2+2M\ell^2=0$, sit at $r_-\approx10\,\mu$m and $r_+\approx80\,\mu$m, the latter at twice the container radius. The requirement is thus $r_+\simeq2M \ll r_{\rm cb}$, which is absolutely not fulfilled. Second, the prescription $g\propto r^{-2}$ anchored at $\tilde r = r_{\rm cb}$ implies $g(r_{\rm cr}) = 1600\,\tilde g$, that is a scattering length $a(r_{\rm cr})\approx3\times10^5a_0\approx17\,\mu$m. This exceeds the transverse oscillator length $l_z=\sqrt{\hbar/m_p\omega_{\rm z}}\approx0.40\,\mu$m by a factor $40$, so the two-dimensional coupling formula used above no longer applies; the associated chemical potential per mass, $c_s^2(r_{\rm cr})$, exceeds $\hbar\omega_{\rm z}/m_p$ by three orders of magnitude, so the cloud is no longer in the transverse ground state; and since the three-body recombination rate scales as $a^4$, the corresponding lifetime collapses from $\sim0.6\,$s at the wall to $\sim10^{-15}\,$s in the core. Third, propagating $U_{\rm cr}$ through the gluing above gives $m_pU_{\rm cr}/k_B\sim4\times10^{-4}\,$K in amplitude, i.e.\ a potential four orders of magnitude deeper than the chemical potential $\mu/k_B \sim 10^{-8}\,$K of the cloud it is supposed to hold, which no in-plane DMD potential can provide. A strongly varying interaction strength is therefore not merely inconvenient but excluded.}

\section{Dynamical blueshift instability for a dynamical mass function}
\label{app:dyn}
In this appendix we give an example from the literature of a dynamical spacetime with a metric analogous to (\ref{eq:ST}), where the mass function is of the Hayward type, but with a time-dependent $M(v)$ where $v=t+r$ is an ingoing null coordinate. The function $M(v)$ is chosen to be similar to a top-hat function but with smooth transitions between the $M=0$ and the $M=const$ regimes \cite{Cardoso_2023}. We suppose that, due to a certain degeneracy between $\ell$ and $M$ in simple UCO geometries like the Hayward model, a similar configuration can be achieved with a time-varying $\ell$ as well. This leads to a spacetime with an inner and outer apparent horizon forming and disappearing in finite time. The former suffers from a finite, but very large blueshift amplification of scalar perturbations. To observers at infinity this could look like a very high-energy jet or bolt.

There is a caveat to this naive generalization to the non-stationary case. Its realization requires a time-dependent external potential, which would lead to a time-dependent flow. This would require a completely different analysis that takes into account $\partial_t\phi$ term in the Bernoulli eq. (\ref{eq:bern}). In the experimental realization, this caveat could be circumvented by imposing an almost precise top-hat variation of $M(v)$ or $\ell(v)$, making the transitions between the value zero and the constant value as discontinuous as possible, given the limitation set by the healing length in a condensate. Such a realization would amount to an effective turning on and off of the trapping (supersonicity). The two regimes (trapping on and trapping off) are both separately stationary, with the time dependence reduced to almost-delta spikes that are not treated.

In BECs, instead of an inflow with a sink in the center and a refilling at the boundary, an oscillatory configuration where trapping is turned on and off may be more viable experimentally. This might be especially true if the experimental setup is confined not only to a disc, but to a ring, effectively yielding a 1+1 dimensional geometry for propagating modes, like in \cite{Garay_2000, Garay_2001}. If this is the case, such dynamical (or almost step-wise stationary) situations might be more realistic to simulate in the BEC analog platform than the fully stationary case in the bulk of this paper and should be kept in mind.

\bibliography{apssamp}

\end{document}